\documentclass[fleqn,usenatbib]{mnras}

\usepackage{newtxtext,newtxmath}

\usepackage[T1]{fontenc}

\DeclareRobustCommand{\VAN}[3]{#2}
\let\VANthebibliography\thebibliography
\def\thebibliography{\DeclareRobustCommand{\VAN}[3]{##3}\VANthebibliography}

\usepackage[table,natural,dvipsnames]{xcolor}
\usepackage{graphicx}
\usepackage{hyperref}
\hypersetup{colorlinks=true,allcolors=blue,pdfborder={0 0 0}}

\newcommand{\Ndiscs}[1]{140} 

\title[Systematic determination of dust properties]{Systematic determination of dust properties for a sample of 133 spatially resolved debris discs}

\author[J. P. Marshall, et al.]{
J. P. Marshall$^{1}$\thanks{Corresponding author, e-mail: jmarshall@asiaa.sinica.edu.tw (JPM)},
S. Hengst$^{2}$,
R. Young$^{3}$,
F. Kemper$^{4,5,6}$,
L. Matr\`a$^{7}$,
N. Pawellek$^{8,9}$,
H. Kobayashi$^{10}$,
\newauthor
P. Scicluna$^{11}$,
S. T. Zeegers$^{12}$
\\
$^{1}$Academia Sinica Institute of Astronomy and Astrophysics, 11F of AS/NTU Astronomy-Mathematics Building, No.1, Sect. 4, Roosevelt Rd, Taipei 106319, Taiwan\\
$^{2}$Centre for Astrophysics, University of Southern Queensland, Toowoomba, QLD 4350, Australia\\
$^{3}$Vanderbilt University, Nashville, Tennessee, United States\\
$^{4}$Institut de Ci\`encies de l'Espai (ICE, CSIC), Can Magrans, s/n, E-08193 Cerdanyola del Vall\`es, Barcelona, Spain\\
$^{5}$ICREA, Pg. Llu\'{\i}s Companys 23, E-08010 Barcelona, Spain\\
$^{6}$Institut d'Estudis Espacials de Catalunya (IEEC), E- 08860 Castelldefels, Barcelona, Spain\\
$^{7}$School of Physics, Trinity College Dublin, the University of Dublin, College Green, Dublin 2, Ireland\\
$^{8}$Institut f\"ur Astrophysik, Universit\"at Wien, T\"urkenschanzstraße 17, 1180 Vienna, Austria\\
$^{9}$Konkoly Observatory, Research Centre for Astronomy and Earth Sciences, E\"otv\"os Lor\'and Research Network (ELKH), Konkoly-Thege Mikl\'os \'ut 15-17,\\ 1121 Budapest, Hungary\\
$^{10}$Department of Physics, Nagoya University, Nagoya, Aichi 464-8602, Japan\\
$^{11}$Centre for Astrophysics Research, University of Hertfordshire, Hatfield, UK\\
$^{12}$European Space Agency / ESTEC Keplerlaan 1, 2201 AZ, Noordwijk, The Netherlands}
\date{Accepted XXX. Received YYY; in original form ZZZ}

\pubyear{2025}

\begin{document}
\label{firstpage}
\pagerange{\pageref{firstpage}--\pageref{lastpage}}
\maketitle

\begin{abstract}
Determination of the composition and size distribution of dust grains in debris discs is strongly dependent on constraining the underlying spatial distribution of that dust through multi-wavelength, spatially resolved imaging spanning near-infrared to millimetre wavelengths. To date, spatially resolved imaging exists for well over a hundred debris disc systems. Simple analytical radiative transfer models of debris dust emission can reveal trends in disc properties as a function of their host stars' luminosities. Here we present such an analysis for 133 debris discs, calculating the dust grain minimum sizes ($s_{\rm min}$), dust masses ($M_{\rm dust}$), and exponents of the size distribution ($q$) in conjunction with their architectures determined at far-infrared or millimetre wavelengths. The distribution of $q$ at far-infrared to millimetre wavelengths is characterised, finding a value of $3.49^{+0.38}_{-0.33}$. We further newly identify a trend between $q$ and $R_{\rm disc}$, which may be indicative of velocity dependent fragmentation, or grain growth at large radii. We find the disc masses inferred from this analysis are consistent with those of protoplanetary discs. Finally, we identify samples of debris discs suitable for further characterisation at millimetre and centimetre wavelengths, expanding the number of spatially resolved systems upon which future studies of these statistics can be based.
\end{abstract}




\section{Introduction}

Planetesimals, i.e. asteroids and comets, form within gas- and dust-rich protoplanetary discs. These minor bodies are the building blocks of planets. Not all planetesimals are swept up into planets during their formation, and remnant belts of planetesimals may persist for the lifetime of the host star and even beyond \citep[see e.g. for recent reviews:][]{2016Farihi,2018Hughes}. Once a high enough threshold velocity is reached, collisions between planetesimals within these belts produces a self-sustaining cascade of smaller bodies down to (sub-)micron sized dust grains. This dusty debris most commonly reveals itself by the presence of excess emission above the stellar photosphere at infrared to millimetre wavelengths. Debris discs may therefore be considered larger, more massive analogues of the Solar system's Asteroid or Edgeworth-Kuiper belts \citep{2020Horner}.

To date, over 1\,000 debris discs have been identified \citep{2023Cao}, but the vast majority of these systems remain spatially unresolved. Knowledge of the underlying dust properties (e.g. grain size distribution, composition, mass, etc.) is degenerate with radial extent (and width) of these discs \citep[e.g.][]{2014aMarshall}. Modelling of debris discs imaged at multiple wavelengths weakens that inherent degeneracy \citep[e.g.][]{1994Sylvester,2011Ertel,2017Hengst,2020Hengst}. Dust emission at far-infrared and (sub-)millimetre wavelengths is sensitive to large dust grains that are unperturbed by radiation forces \citep{2010Krivov}. These grains therefore trace the location(s) of the dust-producing planetesimal belt(s) within a system \citep{2019Pawellek}.

The minimum size of dust grains, $s_{\rm min}$, in the dust size distribution and their relation to the stellar parameters for a sample of spatially resolved debris discs from \textit{Herschel} observations were investigated in \cite{2014Pawellek} and \cite{2015Pawellek}. They found $s_{\rm min}$ consistent with being at, or a few times larger, than the blowout grain size $s_{\rm blow}$ \citep[the size for which dust grains will be ejected from a system by stellar radiation pressure, see e.g.][]{2010Krivov}, for all stars in the sample, consistent with collisional evolution of these discs. They further found no strong dependence of the relation between on $s_{\rm min}$ vs $s_{\rm blow}$ with grain composition or porosity. The absence of discs around lower luminosity stars with $s_{\rm min}$ close to $s_{\rm blow}$ was attributed to greater dynamical stirring in the discs of more luminous stars.

The size distribution of dust grains within a debris disc is often modelled as a continuous power law extending between minimum and maximum grain sizes with a differential form $dn \propto s^{-q} ds$. The size distribution exponent, $q$ has been determined observationally either by fitting the slope of the sub-millimetre emission, or full radiative transfer modelling of the disc spectral energy distribution (SED). At mid-infrared wavelengths \textit{Spitzer} detections of debris discs with a 10~$\mu$m silicate feature \citep{2015Mittal} revealed a broad range of $q$ values spanning 3.5 to 5 and peaking at 4. millimetre wavelengths a small sample of 22 discs reveals a much flatter $q = 3.31~\pm~0.07$ \citep{2016Macgregor,2017Marshall,2021Norfolk}. The value of $q$ is oft assumed to be 3.5, consistent with a steady-state collisional cascade \citep{1969Dohnanyi,1996Tanaka}. Numerical models of dust distributions in collisional equilibrium converge on a slightly steeper value of $q = 3.65$ \citep{2012Gaspar}. Refinements to the Dohnanyi model including a size-dependent tensile strength of the colliding bodies and velocity-dependent fragmentation extends the potential range of $q$ from 3 to 4 \citep[e.g.][]{2003OBrienGreenberg,2010Kobayashi}. Collisional debris from more weakly bound `rubble pile' bodies produces lower values of $q$ \citep{2005PanSari,2012PanSchlichting}, whereas fragmentation of monolithic bodies produces higher $q$ values \citep{1999BenzAsphaug,2012PanSchlichting}. The dust size distribution deviates from a single power law due to the existence of a finite cut-off grain size at $s_{\rm min}$ introducing a `wavy' pattern \citep{2006Krivov,2007Thebault,2019ThebaultKral}. Inclusion of dust removal processes (e.g. Poynting-Robertson drag, stellar winds) can further alter the size distribution for the restricted range of grain sizes where those mechanisms are dominant \citep{2011Wyatt}. These waves are small in amplitude and have yet to be detected observationally. Degeneracies exist in the determination of $q$ from radiative transfer modelling, most importantly with the assumed dust composition \citep{2020Loehne}, which for most debris discs is unconstrained through the absence of solid state mid-infrared spectral features \citep[e.g.][]{2014Chen}.

The dust mass of a debris disc can be inferred from millimetre wavelength emission, assuming the disc to be completely optically thin, and adopting a dust mass opacity \citep[e.g.][]{2017Holland}. The inferred dust mass for debris discs is seen to evolve over time from high masses at younger ages to lower masses at later ages, consistent with the collisional evolution of a parent planetesimal population grinding down into dust \citep{2008Loehne,2022Najita}. Extrapolation of dust masses to total (planetesimal) masses for many of the brightest known debris discs are often in excess of what would be considered plausible for their initial protoplanetary disc masses \citep{2021KrivovWyatt}.

In this work we present an analysis for 133 systems out of an initial sample of 140 spatially resolved debris discs identified in surveys by \textit{Herschel} \citep{2021Marshall,2025Heras}, the Sub-Millimeter Array and the Atacama Large Millimeter/sub-millimeter Array \citep{2016Steele,2025Matra}. A large degree of overlap exists between the samples studied by these facilities such that many, but not all, spatially resolved discs have been observed by both. These surveys have revealed a rich variety of non-axisymmetric structures such as star-disc offsets, warps, and blobs. However, we do not concern ourselves here with accounting for the specifics of every disc architecture in this work. Our goal is to provide a uniform analysis of a large sample of spatially resolved discs to determine the properties of debris dust as a population rather than precise modelling of every individual system.

The remainder of the paper is laid out as follows. We present a summary of the underlying observations and the methods used to model the compiled data sets in Section \ref{sec:obs}. The modelling and analysis are presented in Section \ref{sec:mod}. A summary of the dust properties and their trends are given in Section \ref{sec:res}. We compare our findings to previous studies of the dust grain properties of debris discs in Section \ref{sec:dis} before presenting our conclusions in Section \ref{sec:con}.

\section{Observations}
\label{sec:obs}

The sample of discs underlying this study was assembled from literature studies of debris discs that have been spatially resolved in thermal emission. We focus on the analyses of spatially resolved discs with \textit{Herschel} published in \citet{2021Marshall} and \citet{2025Heras}, representing the largest such homogeneous sample, and at millimetre wavelengths the REASONS (ALMA and SMA) sample of \citet{2025Matra} along with additional systems from \cite{2016Steele}. For these studies the disc architectures were modelled in a consistent fashion facilitating their combination and comparison. We do not include discs that have only been resolved in scattered light \citep[e.g.][]{2020Esposito}, or young debris discs in star forming regions \citep[e.g.][]{2016LiemanSifry} that lack sufficient photometry at infrared wavelengths to model the dust emission. There were \Ndiscs\, individual systems within the combined data set that formed our initial sample, with 74 systems with architectures coming from ALMA or SMA observations and 66 systems with resolved extents based on \textit{Herschel} observations.

The photometric data and stellar atmosphere models assembled for each system were taken from the stellar database (SDB)\footnote{\href{http://drgmk.com/sdb/}{http://drgmk.com/sdb/}}, for which the methods are presented in \cite{2019Yelverton}. The photometry used in determination of the infrared excess predominantly come from observations by \textit{WISE} \citep{2010Wright}, \textit{Akari} \citep{2010Ishihara}, \textit{Spitzer} \citep{2006Su,2009Bryden,2014Chen}, \textit{Herschel} \citep{2013Eiroa,2014Thureau,2016Montesinos,2018Kennedy}, JCMT/SCUBA-2 \citep{2017Holland}, the SMA and ALMA \citep{2025Matra,2016Steele}.

A warmer, inner component was identified around 33 debris disc systems in the sample\footnote{These systems are identified in Table \ref{tab:app_results}.}. These were identified using SED fitting of modified blackbodies to the excess emission and {\sc multinest} to discriminate between one and two component models for the debris discs \citep{2019Yelverton}. The contribution of a warm component to the total emission of a system was subtracted from the measured fluxes prior to radiative transfer modelling so as not to bias the results of our analysis. We complement the data taken from the SDB with additional photometry, predominantly at millimetre wavelengths, which were absent from the compiled data sets within the repository \citep[e.g. HD~16743, HD 138965;][]{2023bMarshall,2025Marshall}. 

\section{Modelling and analysis}
\label{sec:mod}

Modelling the dust emission of the debris disc sample was undertaken using the Python-based 1D analytical radiative transfer code {\sc artefact}\footnote{\href{https://github.com/jontymarshall/artefact}{https://github.com/jontymarshall/artefact}} \citep{2023bMarshall}. As inputs to the code we took the radius and width of the disc from the imaging observations, and the compiled photometry, stellar parameters, and photosphere model from SDB. 

For the disc architectures, each system was modelled as a Gaussian ring, with an architecture defined by a flux density $f_{\rm disc}$, peak radius $R_{\rm disc}$, fractional width $\Delta R_{\rm disc}$, inclination $i$, and position angle $\phi$. When assigning an architecture for each system, we preferentially selected the architecture derived from ALMA as having the best combination of angular resolution and signal-to-noise, followed by the SMA, and lastly \textit{Herschel}. The sole exception to this procedure was HD~48682, for which the SMA detection is marginal and the \textit{Herschel} data are therefore a better constraint on the disc architecture \citep{2020Hengst}.

For the dust composition we used the complex optical constants $n$ and $k$ of astronomical silicate \citep{2003Draine}, with dust optical properties calculated by Mie theory using the Python package {\sc MiePython}\footnote{\href{https://github.com/scottprahl/miepython}{https://github.com/scottprahl/miepython}} \citep{MiePython}. The total sample of discs was split into two parts on the basis of the availability of (sub-)millimetre photometry beyond 450~$\mu$m to constrain the particle size distribution. For those systems with sufficient data, we fitted the minimum grain size $s_{\rm min}$ (with a fixed $s_{\rm max} = 3$~mm), exponent of the size distribution $q$, and dust mass $M_{\rm dust}$. For systems without sufficient data at (sub-)millimetre wavelengths to constrain the size distribution exponent $q$ was held fixed as 3.5 in the modelling. We model the SEDs of all systems out to 10~mm for comparison with existing long wavelength measurements \citep{2016Macgregor,2021Norfolk}.

The parameter space was explored using {\sc emcee} \citep{2013ForemanMackey}. For each system we initialise 50 walkers with randomly generated, uniformly distributed values of $\log_{10} s_{\rm min}$ between -1 and 1 (in $\mu$m), $\log_{10} M_{\rm dust}$ between -6 and -3 (in $M_{\oplus}$), and $q$ between 3.8 and 3.2. The bounds on the parameters were $\log_{10} s_{\rm min}$ between -2 and 2, $\log_{10} M_{\rm dust}$ between -9 and 0, and $q$ between 2 and 5. For each realisation of the model a least squares fit was made to the observations and passed to {\sc emcee} as the log probability. We ran the walkers in chains with a burn-in of 100 steps and thereafter until all parameters had passed a threshold of 50 auto-correlation times, which was typically achieved in around 2\,000 steps. The posterior probability distribution was assembled using the final 10\,000 realisations of the model. The posterior probability distributions were generally monomodal and normally distributed, although several systems had a pronounced skew to either $s_{\rm min}$ or $q$; these findings will be detailed in the following sections. The final maximum amplitude probability model parameters for each disc ($s_{\rm min}$, $M_{\rm dust}$, and $q$ where possible) and their associated uncertainties were determined using the 16$^{\rm th}$, 50$^{\rm th}$, and 84$^{\rm th}$ percentiles of the distribution.

\section{Results}
\label{sec:res}

Here we present the statistics and trends for the three dust properties, i.e. minimum grain size, exponent of the size distribution, and dust mass, considered in this analysis. These were compared against system properties, such as the stellar luminosity and disc radius, to search for trends in their behaviour. Some of these trends have been previously identified, such as the relationship between minimum grain size and stellar luminosity \citep{2014Pawellek}, but the expanded sample size presented here makes revisiting these trends worthwhile. A complete summary of the results of the radiative transfer modelling are provided in tabulated form in Appendix \ref{app:results}.

\subsection{Anomalous SEDs}

The first result we must note is that seven of the \Ndiscs\, systems modelled in this work have SEDs that do not adequately match the observations. Six of the systems, HD~25821, HD~32977, HD~36546, HD~113766, HD~158643, and HD~278932, are distinct from the bulk of the modelled systems through the very small minimum grain size ($\leq$ 0.2~$\mu$m) derived in the modelling. These systems have poorly sampled SEDs with only one or two far-infrared data points constraining the cold excess and evidence of substantial mid-infrared excess in \textit{WISE} W3 and/or W4 bands. The far-infrared emission for those sources being poorly constrained means that modelling cannot meaningfully constrain the dust properties of the debris disc. Exhibiting substantial mid-infrared excess suggests these systems could be younger, less-evolved discs. The radiative transfer code used for this analysis is unsuitable to model such systems and they have therefore been omitted from further analysis. 

The final system, HD~90089, has a relatively cold excess (peaking at 160~$\mu$m), similar to the `cold debris discs' identified by Herschel DUNES \citep{2011Eiroa,2013Marshall,2013Krivov}. Radiative transfer modelling reveals the spatially resolved extent of HD~90089's disc is inconsistent with the low temperature of the excess emission, from which we may infer the excess may instead be due to contamination from a background galaxy. Omitting all of these unusual systems from the total sample leaves a population of 133 debris discs which we examine in the remainder of this section.

\subsection{Minimum grain size}

\begin{figure}
    \centering
    \includegraphics[width=\columnwidth]{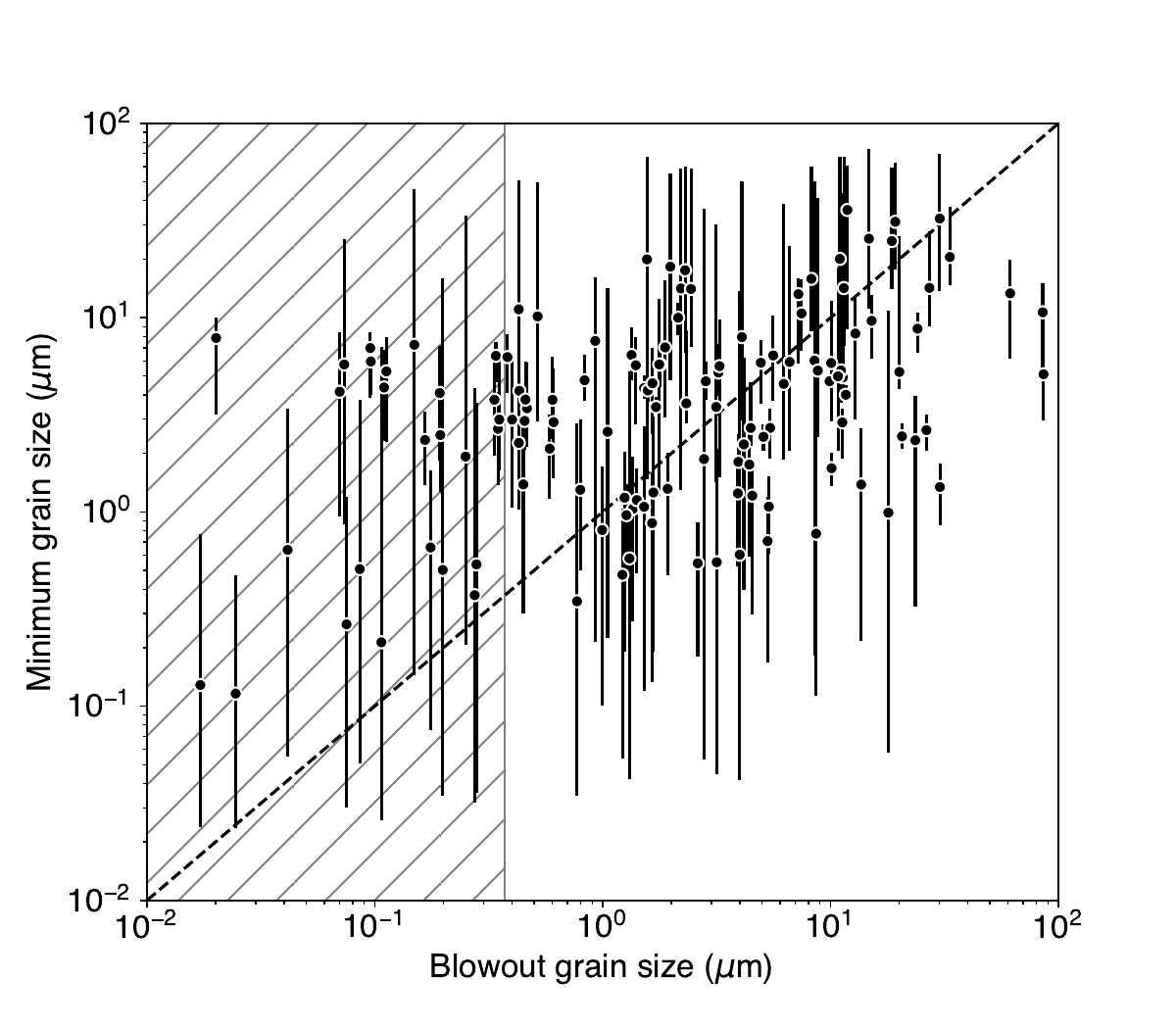}
    \caption{Plot of $s_{\rm min}$ vs. $s_{\rm blow}$. The dashed line denotes equivalence between $s_{\rm min}$ and $s_{\rm blow}$, whilst the hatched region denotes the range of calculated blowout grain sizes that are not meaningful as the associated stellar luminosity is too small to produce high $\beta$ grains. It is clear that for most systems, particularly around high luminosity stars, a substantial contribution to their excess emission comes from sub-blow out grains.}
    \label{fig:smin_vs_sblow}
\end{figure}

The relationship between the calculated minimum grain size, $s_{\rm min}$, and the blowout grain size, $s_{\rm blow}$, for each system (a function of stellar luminosity) is shown in Figure \ref{fig:smin_vs_sblow}. A clear trend of increasing minimum grain size  as a function of stellar luminosity is seen in the sample. We model the sample distribution with a power law following \cite{2014Pawellek} using an error-weighted least squares fit obtaining a slope and intercept of 0.21 and 2.1~$\mu$m, respectively. This is consistent with the previous determination based on the smaller sample used by \citet{2014Pawellek}, who found slope and intercept values of 0.20~$\pm$~0.07 and 3.2~$\pm$~0.7, and statistically significant with a Spearman correlation coefficient of 0.29 and a $p$-value of 0.007. Below 1.05~$L_{\odot}$ we can calculate that the stellar radiation pressure will not be sufficient to remove dust grains from the system \citep{1979Burns}, therefore the blowout grain sizes for those low luminosity are not meaningful. The equivalent blowout grain size for 1.05~$L_{\odot}$ is 0.37~$\mu$m and is denoted by the hatched region in Figure \ref{fig:smin_vs_sblow}. At intermediate stellar luminosities, a mixture of systems with minimum dust grain sizes both above and below the blowout radius is present in the sample. Above 20~$L_{\odot}$ all of the debris discs have a minimum grain size smaller than the blowout radius, such that the dust emission observed at infrared wavelengths comes predominantly from transient dust grains \citep{2019ThebaultKral}.

For a minority of systems the upper bound of the dust grain size is poorly constrained (i.e. the positive error bar on $s_{\rm min}$ is large). This is a natural consequence of modelling sparsely sampled SEDs. The systems for which this is true are predominantly those for which we have weak, or non-existent constraints on the sub-millimetre slope, particularly bridging the gap between far-infrared and millimetre photometry. Improving these results will require a dedicated campaign of observation to fill in the required observations.

\begin{figure}
    \centering
    \includegraphics[width=\columnwidth]{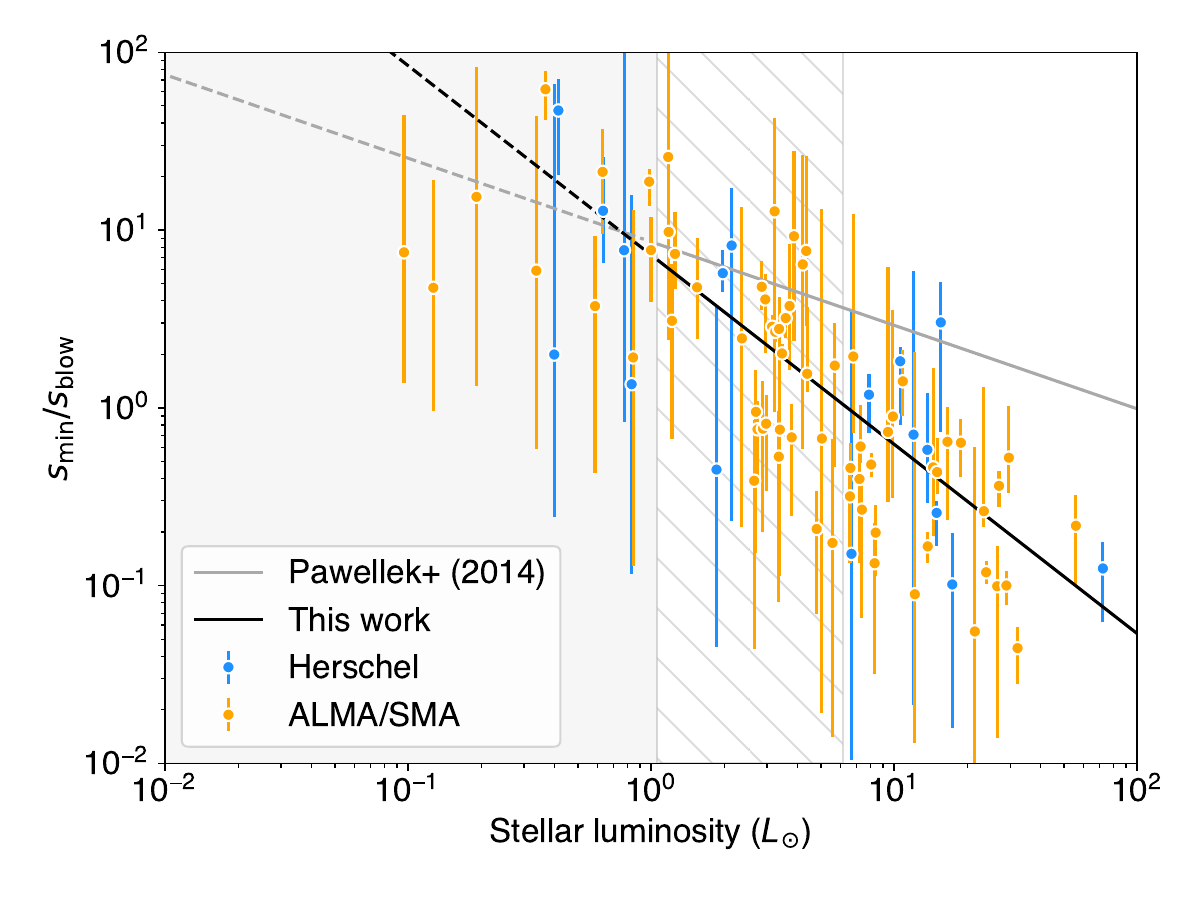}
    \includegraphics[width=\columnwidth]{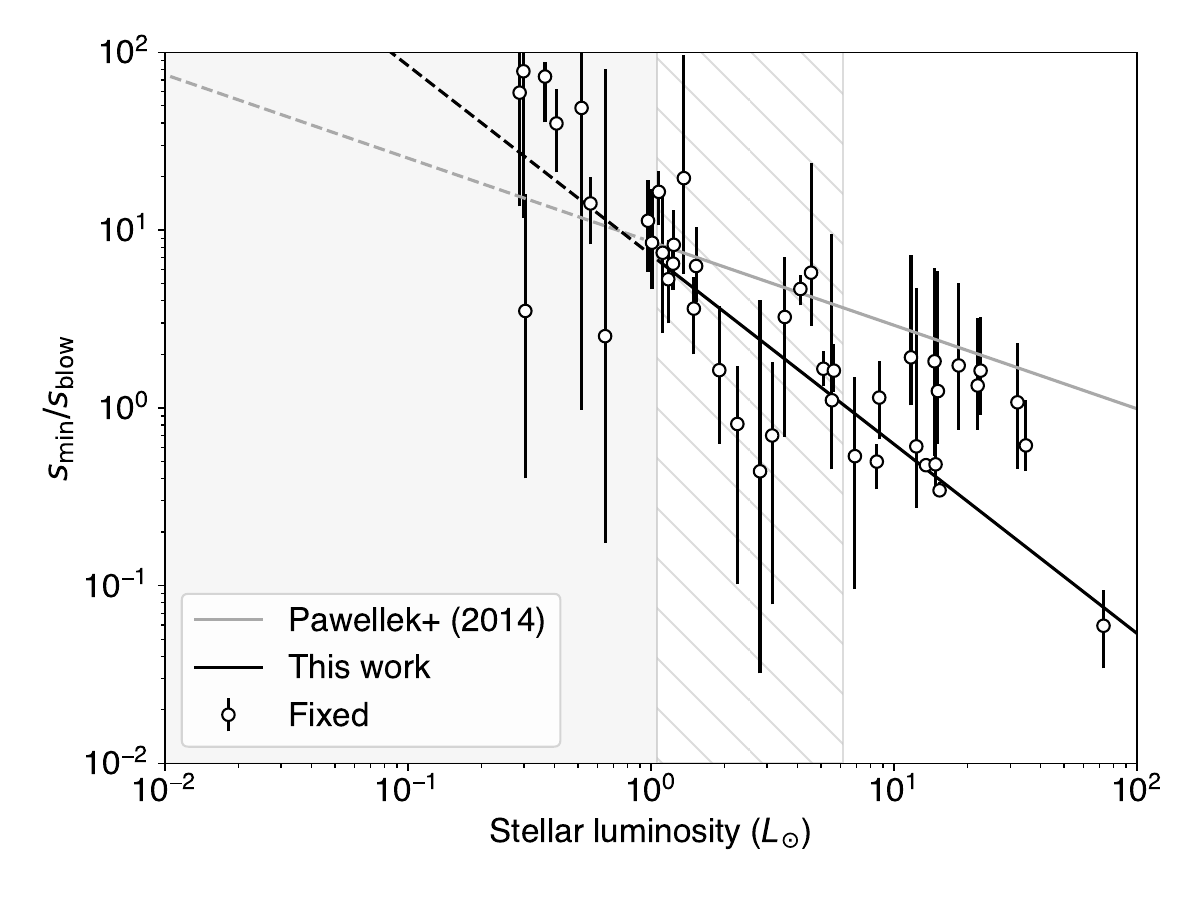}
    \caption{\textit{Top}: Plot of $s_{\rm min}/s_{\rm blow}$ vs. $L_{\star}$. Blue data points are for \textit{Herschel}-resolved architectures, whereas orange data points are for ALMA- or SMA-resolved architectures. Uncertainties are 1-$\sigma$. The shaded region denotes stellar luminosities for which no blowout grain size exists and the hatched region denotes stellar luminosities for which a range of grain sizes are blown out. The two straight lines denote the relationship between $s_{\rm min}/s_{\rm blow}$ determined by \citet{2014Pawellek} (grey line) and the fit to the ensemble presented here (black line). \textit{Bottom}: Same as above, except the white data points denote discs with $q = 3.5$.}
    \label{fig:sratio_vs_lstar}
\end{figure}

In Figure \ref{fig:sratio_vs_lstar} we present the ratio of $s_{\rm min}$/$s_{\rm blow}$ as a function of stellar luminosity and the exponent of the size distribution. In both panels, shaded regions denote the regions of stellar luminosity below which there is no blowout radius for astronomical silicate dust grains ($L_{\star} \leq 1.05~L_{\odot}$) and below which there is a range of dust grains removed by radiation pressure ($1.05~L_{\odot} < L_{\star} \leq 6.15~L_{\odot}$). Above $6.15~L_{\odot}$ there is a singular minimum bound grain size for each system. Uncertainties on $s_{\rm min}$ are largest for discs with weak constraints on their sub-millimetre emission, lacking photometric data between 160 and 850~$\mu$m. There is a clear trend between the stellar luminosity and the ratio $s_{\rm min}/s_{\rm blow}$, albeit with substantial scatter and large intrinsic uncertainties in the determination of $s_{\rm min}$. A linear regression fit using the {\sc SciPy} function \textit{linregress} finds a Spearman correlation coefficient of -0.69 and a $p$-value 10$^{-16}$, demonstrating the sample has a distribution strongly deviating from a flat slope. Comparison between the modelling presented here and previous results from \cite{2014Pawellek} and \cite{2015Pawellek} studying a smaller sample of 32 discs (all of which are included in this analysis). The dependence of $s_{\rm min}/s_{\rm blow}$ is much steeper here than the trend for astronomical silicate dust presented in \cite{2015Pawellek} (shown in Figure \ref{fig:sratio_vs_lstar}). Their sample was smaller, lacked the high resolution architectures at millimetre wavelengths, and had no constraints on the disc widths. The higher angular resolution data result in larger inferred disc radii for the systems in this sample, such that $s_{\rm min}$ is decreased and the overall slope of the distribution becomes steeper.

\subsection{Exponent of the dust size distribution}

Out of the 133 systems, 96 have sufficient photometric coverage at (sub-)millimetre wavelengths to make a determination of the dust grain size distribution exponent $q$ from SED modelling. Debris discs typically have values of $q$ between 3 and 4. All of the systems modelled here have measured exponents within the range 2 to 5 with 89 systems having measured values between 3 and 4. Typical uncertainty in the measurement of $q$ is $\pm0.25$ for the sample. Outliers with much larger uncertainties in $q$ (up to $\pm 1.0$) are present in the results, and those systems are characterised by having only photometry shortward of 450~$\mu$m constraining their sub-millimetre slopes -- sufficient to model $q$ as a free parameter, but only enough to weakly constrain it.

The discs with the steepest exponents are HD~113556, HD~158633, and HD~159492, for all of which $q > 4$. The sharp drop off in their sub-millimetre SEDs are reminiscent of the `steep SED' discs identified at far-infrared wavelengths by \cite{2012Ertel}. However, the large uncertainties leave the range of possible $q$ values consistent with typical debris discs for all three systems, leaving their true nature ambiguous. At the flatter end of $q$ values, HD~27290, HD~48370, HD~109085, and HD~131488 all have $q < 3$. Again, the associated uncertainties mean the calculated values for these systems are not significantly above the expected range for $q$. All four systems exhibit observational properties consistent with a large contribution from small (warm) dust grains exhibiting strong mid-infrared excesses \citep{2013Broekhoven-Fiene,2016Moor,2024Pawellek}, and HD~109085 in particular having dust spectral features associated with a giant collision event \citep{2012Lisse}. 

\begin{figure}
    \centering
    \includegraphics[width=\columnwidth]{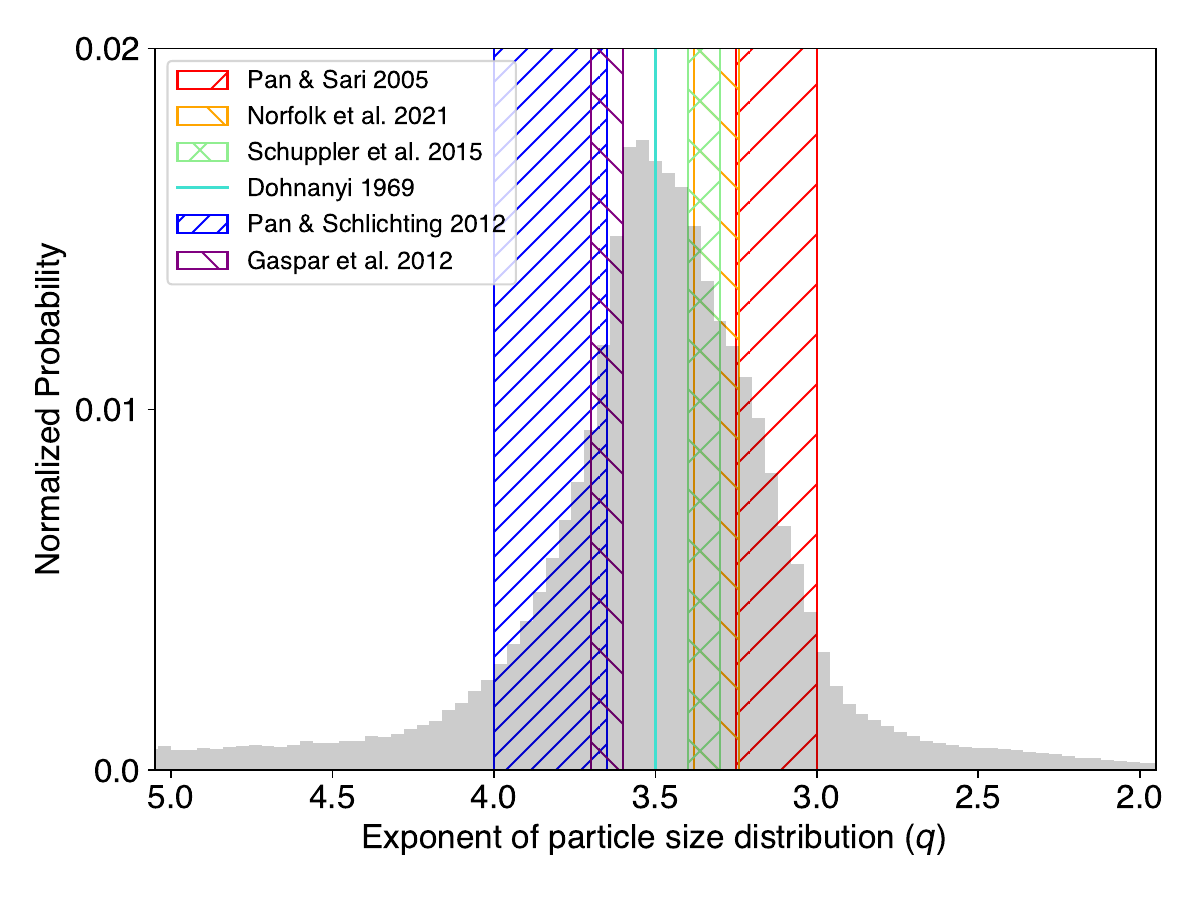}
    \caption{Histogram of $q$ values derived from modelling. The distribution is assembled from the posterior probability distributions of individual systems. Hatched regions on the plot denote the regions associated with $q$ expected for different collisional models (see text for details).}
    \label{fig:q_hist}
\end{figure}

The overall distribution of $q$ is presented in Figure \ref{fig:q_hist}, and is derived from the posterior probability distributions of the individual fits. Based on the probability distribution the median value of $q$ is 3.49$^{+0.38}_{-0.33}$, consistent with the infinite steady-state collisional cascade of \cite{1969Dohnanyi,1996Tanaka}. The distribution can be seen to have a slight skew toward flatter rather than steeper values (compared to a value of 3.5), but it is not significantly skewed. We compare the measured values of $q$ in Figure \ref{fig:q_hist} to various theoretical \citep{2005PanSari,2012PanSchlichting,2013Gaspar} and observational results \citep{2015Schuppler,2021Norfolk}. The theoretical modelling can reproduce values of $q$ in the range 3 to 4, depending on the assumptions. \cite{2003OBrienGreenberg} demonstrated analytically that a range of $q$ beyond the 3.5 obtained under the assumptions of \citet{1969Dohnanyi} can be obtained by consideration of the tensile strength of the colliding bodies, with weakly bound bodies (`gravity regime') having $q < 3.5$ and strongly bound bodies (`strength regime') having $q > 3.5$. In the presence of a size-dependent velocity distribution and viscous damping of the collision fragments further modifications to $q$ are obtained in the gravity regime \citep[$q \simeq 3$][]{2005PanSari,2012PanSchlichting}, and strength regime \citep[$q > 3.65$][]{2012PanSchlichting}. The expected range of $q$ is further dependent on the relative velocities of the bodies involved in the collision. In \citet{2012Gaspar}, good consistency for a broad range of disc models in collisional equilibrium is found, converging to a value of $q = 3.65~\pm~0.05$, with some small dependence on the disc width and scale height. For comparison between populations we calculate the error-weighted mean of the sample, being $q = 3.45~\pm~0.08$, which is slightly higher but still consistent with the value derived from millimetre-wavelength measurements of $q = 3.31~\pm~0.07$ from \cite{2021Norfolk}, and lower than the distribution of $q$ for mid-infrared discs observed by \textit{Spitzer}, which peaked around 4.0 \citep{2015Mittal}. This suggests an evolution in the measured value of $q$ as a function of wavelength (contributing grain size), with the far-infrared to millimetre derived slopes being broadly consistent. We note that the value of $q$ at mid-infrared wavelengths is determined from discs with solid-state silicate features such that it is unclear if the inferred value of $q$ is representative of either the small grains producing the spectral features or the bulk of the dust.

Previously there has been some suggestion that the presence of cold molecular CO gas in a debris disc may moderate collisions between dust grains and thereby influence the size distribution \citep{2021Norfolk}. A population of around 20 systems with gas have been identified to date encompassing both molecular gas and atomic line detections \citep{2020Marino}. In this sample there are 11 debris discs with CO gas detections, which are predominantly around young, early-type stars \citep{2017Moor}. The abundance of gas in these systems is such that its origin may be either primordial or generated in planetesimal collisions \citep{2017Kral,2020Marino}. We can compare the calculated values of $q$ for the systems with detected gas in the sample versus the overall distribution to see if they are statistically different. Within the same range of stellar luminosities (6.6 to 24~$L_{\star}$) there are another 30 stars in the sample. Examining the distribution of both $s_{\rm min}$ and $q$ for gas-bearing systems as a function of $L_{\star}$, we find that both are consistent with the distribution of the wider sample.

\begin{figure}
    \centering
    \includegraphics[width=\columnwidth,trim={0 0 0 0},clip]{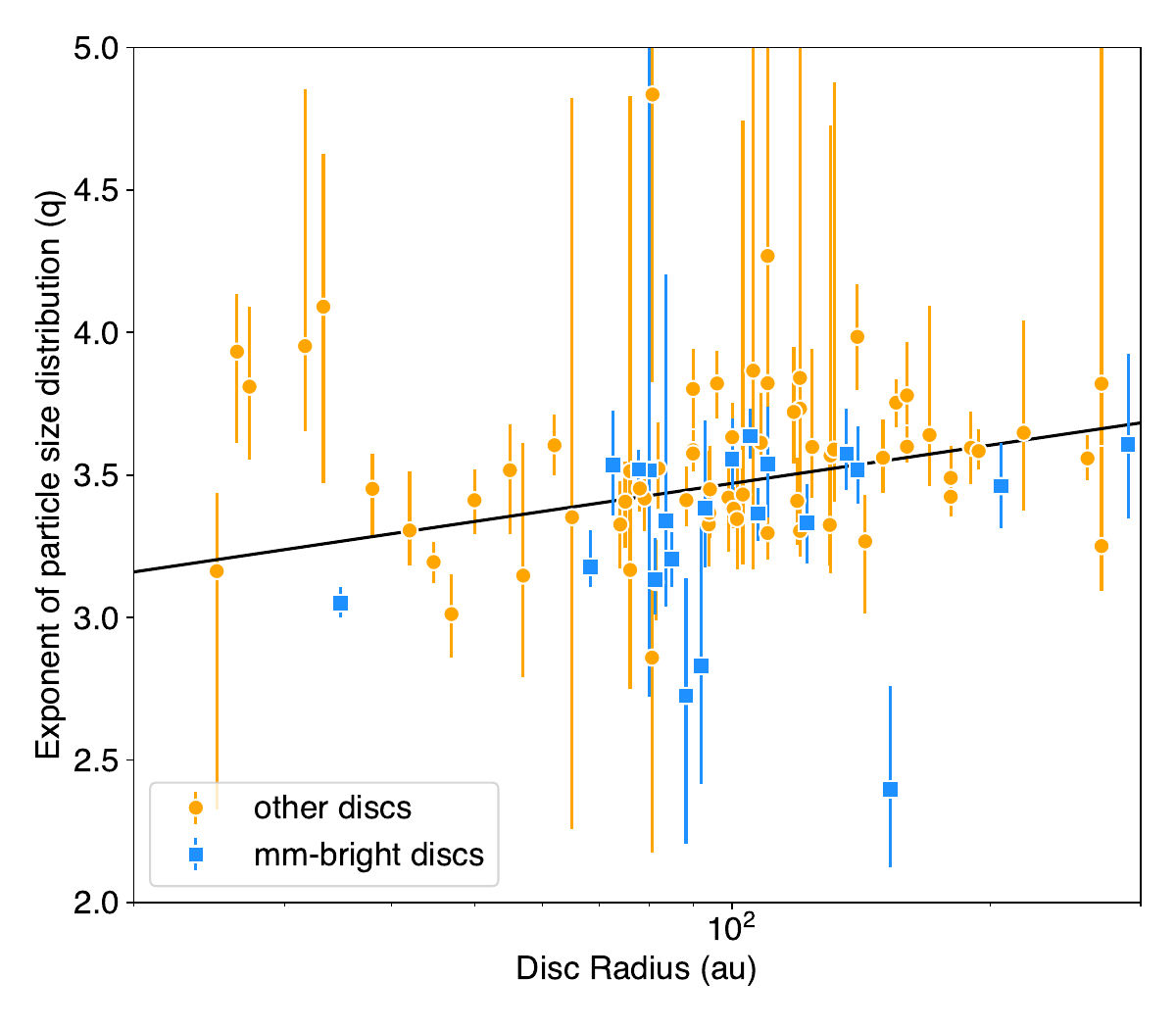}
    \caption{Plot of $q$ vs. $R_{\rm disc}$. Circular data points are for systems for which $q$ could be determined from modelling. Square data points denote systems that have also been detected at 7 or 9~mm (dubbed `mm-bright'). The solid line denotes a power law fit to the observations illustrating a trend between $R_{\rm disc}$ and $q$.}
    \label{fig:q_vs_rdisc}
\end{figure}

We find a trend between $q$ and $R_{\rm disc}$ as shown in Figure \ref{fig:q_vs_rdisc}. We fit the distribution of $q$ vs. $R_{\rm disc}$ with a straight line ($q = m \log_{10}(R_{\rm disc}) + c$) using the {\sc SciPy} \textit{lmfit} routine finding best-fit parameters of $m = 0.49~\pm~0.09$ and $c = 2.48~\pm~0.18$. The significance of this trend can be characterised by the Spearman correlation coefficient ($\rho_{\rm s}$). The statistic $\rho_{\rm s} = 0.26$ and probability $p = 0.013$ were calculated using the {\sc SciPy} {\it stats.spearmanr} function, suggesting a positive correlation exists between $q$ and $R_{\rm disc}$ (or rather, it is highly unlikely we would see the observed distribution if no correlation existed). We test the robustness of this correlation by fitting the trend to the systems with ALMA-derived radii finding values of $\rho_{\rm s} = 0.35$ and $p = 0.005$, such that the significance of the correlation is increased; the majority of sources have ALMA data and these generally have lower uncertainties due to greater precision and denser sampling of the source SEDs.

\begin{figure}
    \centering
    \includegraphics[width=\columnwidth]{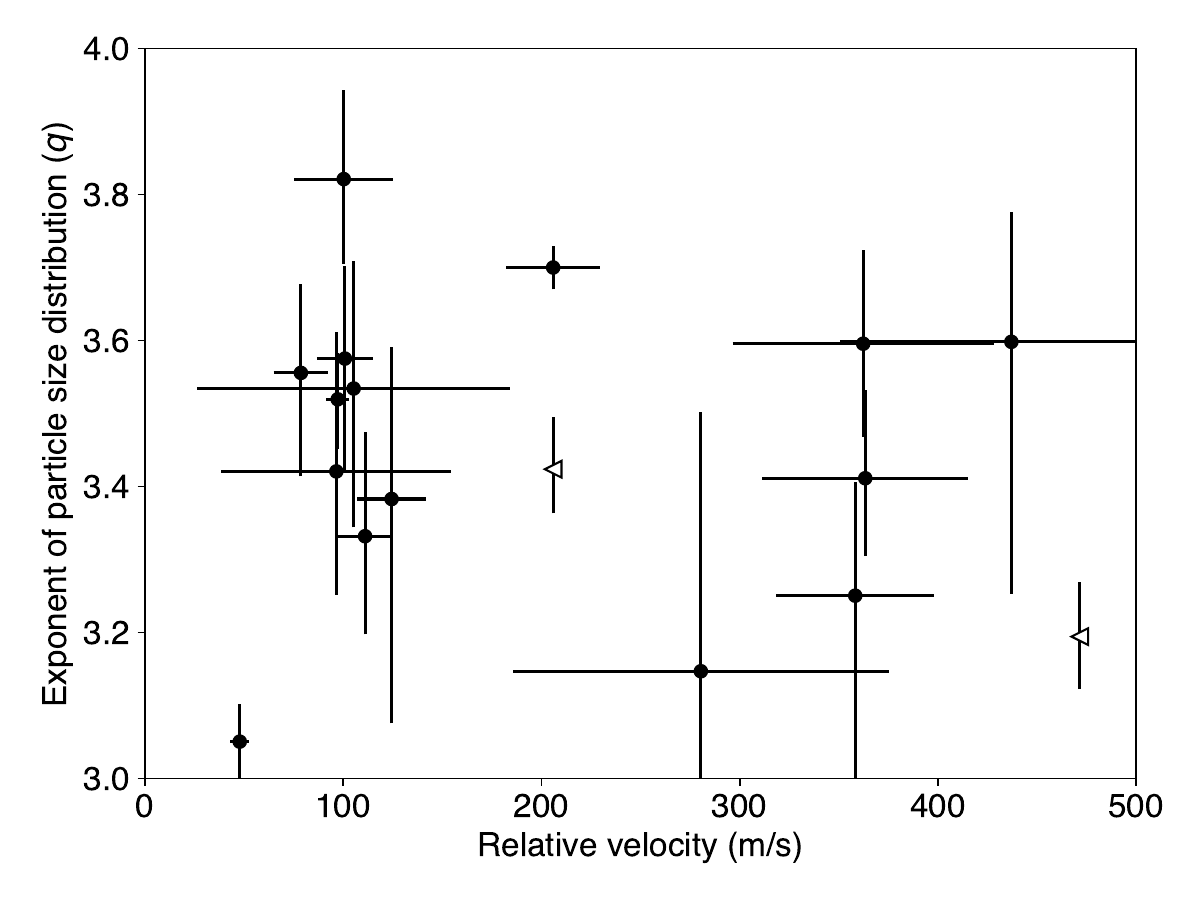}
    \caption{Plot of $q$ vs. $v_{\rm rel}$. Data points are for debris discs with measured $q$ values from this work and the relative velocities are calculated from the disc scale heights in \citet{2023Terrill}. We see no strong trend between $q$ and $v_{\rm rel}$, contrary to what might expected from theoretical modelling \citep{2012Gaspar}, suggesting other factors are contributing to the value of $q$ determined from the observations.}
    \label{fig:q_vs_vrel}
\end{figure}

The parameters of the best-fit trend are consistent with $q \propto \sqrt R_{\rm disc}$ or equivalently $q \propto 1/v_{\rm Kep}$. Modelling by \cite{2012Gaspar} showed that the dust size distribution becomes flatter for higher collisional velocities within debris discs (between 0.1 and 3 km/s), but the Keplerian velocity of a body is not representative of its collisional velocity. A handful of debris disc systems have had their vertical scale heights measured with values of $h$ ranging between 0.02 and 0.21 \citep{2023Terrill}. Assuming the scale height is a proxy for dynamical stirring (i.e. systems with larger $h$ have larger collision velocities), we should see in Figure \ref{fig:q_vs_vrel} that the value of $q$ is anti-correlated with $h$. However, this is not the case due to the intrinsic scatter and the large uncertainties in both $q$ and $v_{\rm rel}$. Alternatively, the relationship could be interpreted as evidence for dust grain growth in the outskirts of larger debris discs where relative impact velocities are lower \citep{2024Kadono}. 

\subsection{Dust mass and disc mass}

\begin{figure}
    \centering
    \includegraphics[width=\columnwidth]{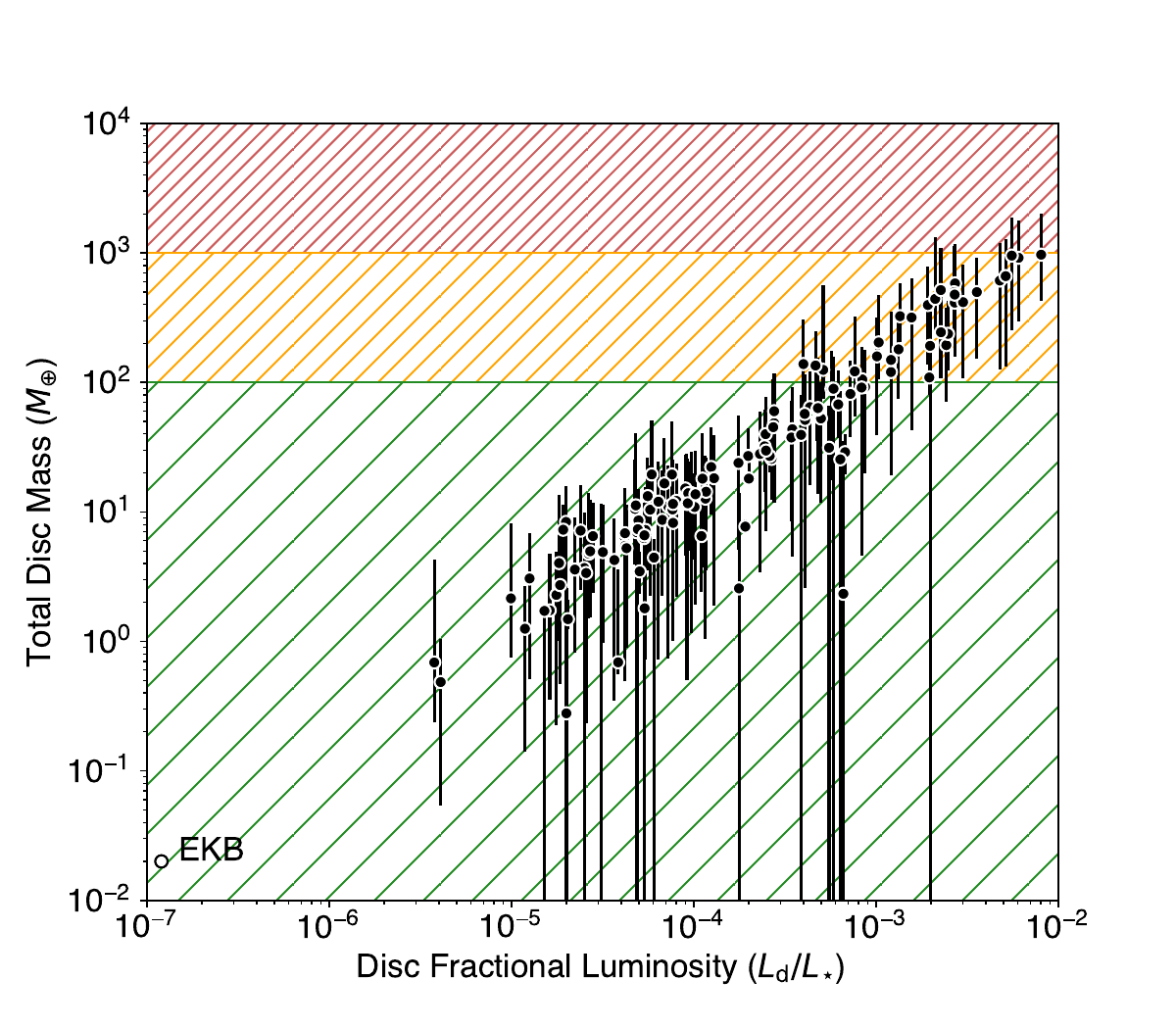}
    \caption{Plot of $M_{\rm disc}$ vs. $L_{\rm d}/L_{\star}$. Black data points denote masses calculated using a three component power law size distribution for bodies up to 400~km in size following \citet{2021KrivovWyatt} with the hatched regions denoting regions for which disc masses were considered marginal or too large based on the same analysis. The white labeled data point denotes the Edgeworth-Kuiper belt.}
    \label{fig:mdisc_vs_fraclum}
\end{figure}

We calculate dust masses for grains up to 3~mm for the debris discs in this work. In Figure \ref{fig:mdisc_vs_fraclum} we present the disc masses for the systems in this sample as a function of the mass of the host star. These masses are calculated using the equations from \cite{2021KrivovWyatt}, assuming power law size distributions of $q_{\rm med} = 3.7$ between 3~mm and 1~km, and $q_{\rm large} = 2.8$ between 1~km and 200~km, the assumed radius of the largest bodies in the planetesimal belt. For the three systems with $q \geq 4$ the disc mass extrapolation from \cite{2021KrivovWyatt} is inappropriate, so we instead adopt a mass opacity and calculate the dust masses for these systems from the measured (or interpolated) flux density at 1.3~mm and then integrate the three power laws across the same size ranges.

Contrary to the results of \citep{2021KrivovWyatt}, we find that none of the disc masses determined from this work are excluded based on the inferred masses of protoplanetary discs.

\begin{figure}
    \centering
    \includegraphics[width=\columnwidth]{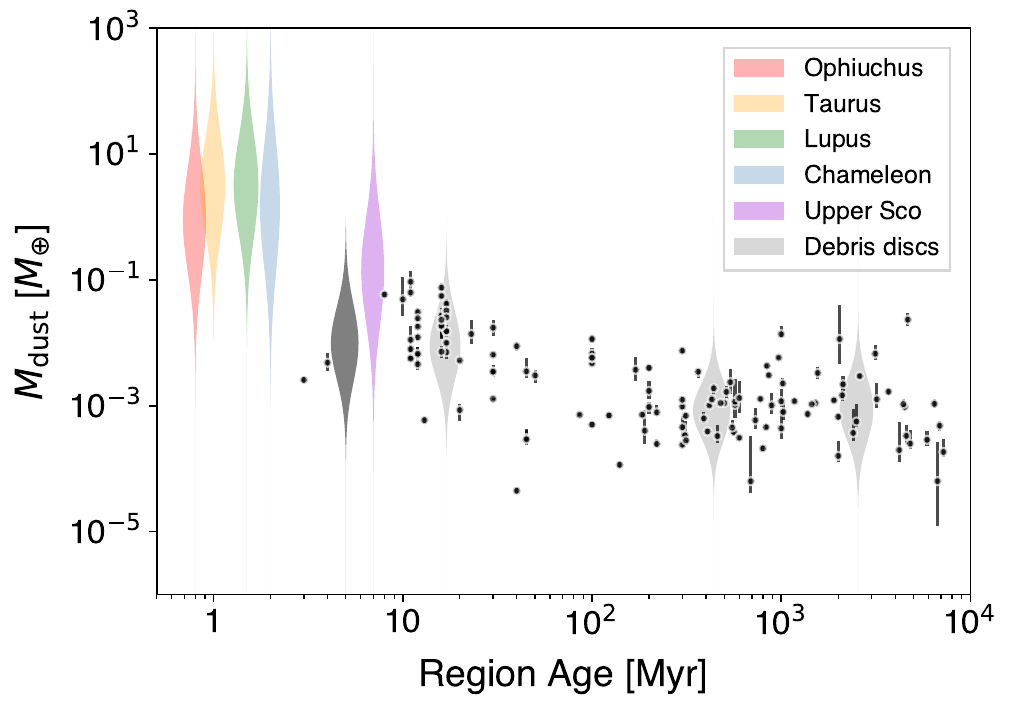}
    \caption{Violin plot of $M_{\rm dust}$ vs. time comparing the dust masses for star forming regions against the distribution of debris discs. Coloured shaded regions show the distributions for the Chameleon, Lupus, Ophiucus, Taurus, and Upper Sco star forming regions, following \citet{2023Manara} (see text for details). The light grey shaded regions show the same distribution for debris disc dust masses for three age ranges, $t =~$ 10--100~Myr, 100--1000~Myr, and 1000-10\,000~Myr, plotted at the median age for each interval. The dark grey shaded region shows the backward extrapolated dust mass of the ensemble for $t = 5~$Myr based on the disc evolution models of \citet{2007Wyatt}. Solid data points show the calculated dust masses for the debris disc sample.}
    \label{fig:mdust_vs_age}
\end{figure}

We also compare the distribution of dust masses for the ensemble to values determined for star forming regions. In Figure \ref{fig:mdust_vs_age} we present the distribution of dust masses for several nearby star forming regions, namely Chameleon \citep{2016Pascucci,2018Long,2021Villenave}, Lupus \citep{2016Ansdell,2016Cleeves,2018Ansdell,2020Sanchis,2019Tsukagoshi}, Ophiucus \citep{2019Williams}, and Taurus \citep{2013Andrews,2014Akeson,2018WardDuong,2019Akeson,2019Long}, and Upper Sco \citep{2014Carpenter,2016Barenfeld,2016vanderPlas}, against the calculated dust masses from this work. The debris disc dust mass distribution at 5~Myr is calculated by backward extrapolation using the stellar ages and observed radii following the model of \citet{2007Wyatt}. Whilst this method is fraught with uncertainty, particularly for older stars ($> 100~$Myrs) whose ages are only weakly constrained, we find that the inferred debris disc dust masses are broadly consistent with the distributions of class II discs in nearby star forming regions.

\subsection{Comparison of predicted and observed millimetre fluxes}

\subsubsection{Band 6 predictions}

\begin{figure*}
    \includegraphics[width=0.48\textwidth]{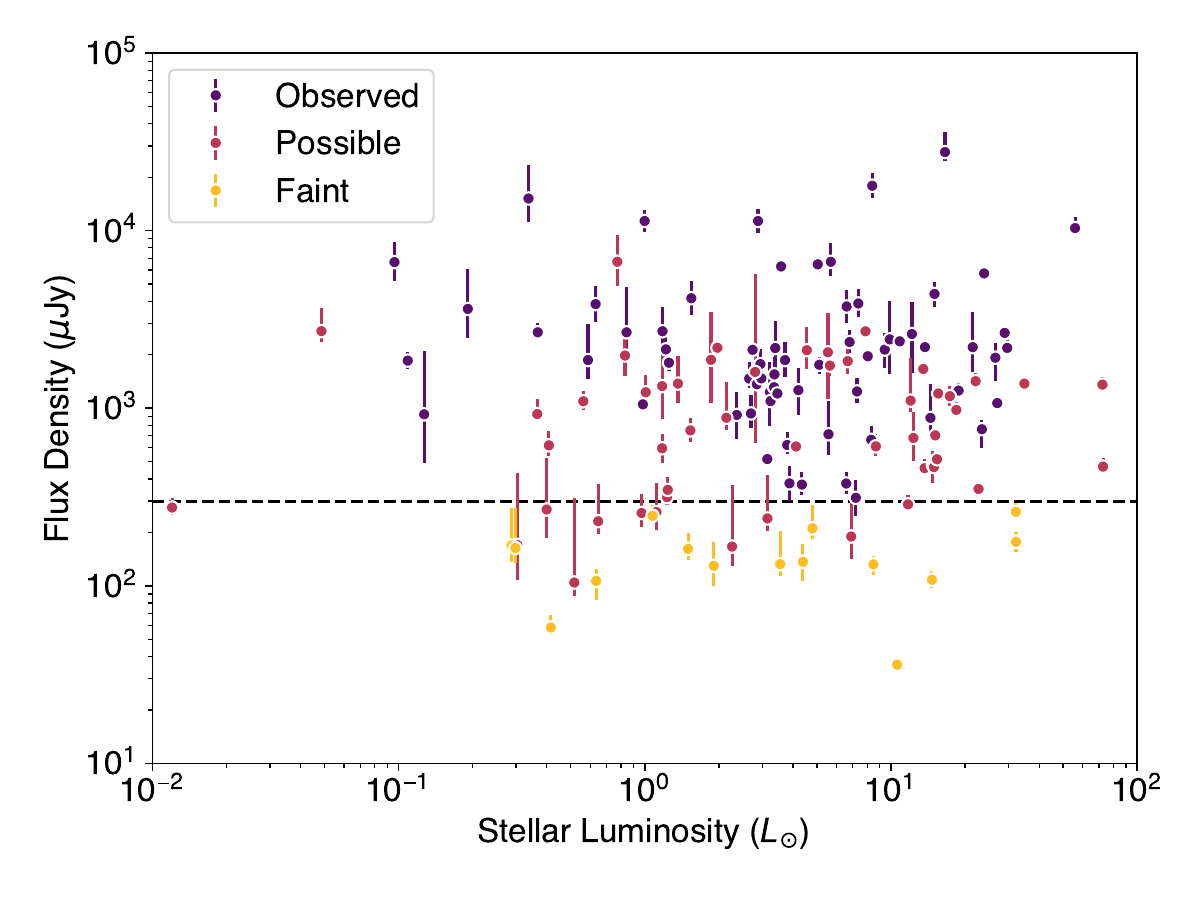}
    \includegraphics[width=0.48\textwidth]{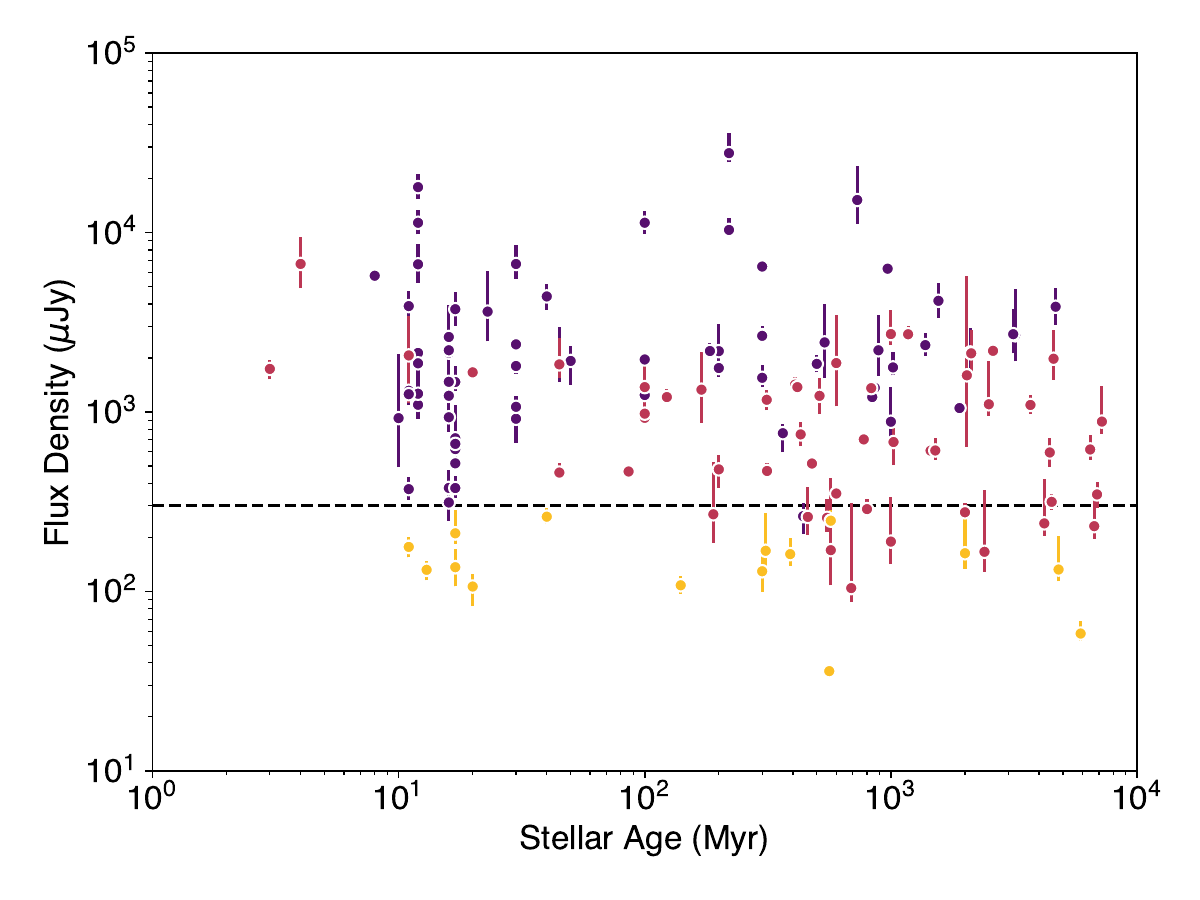}
    \caption{Here we show the potential for additional spatially resolved millimetre wavelength imaging to add to the ensemble of debris discs. The distribution of predicted (and observed) disc flux densities are presented as a function of stellar luminosity (left) and system age (right). Coloured data points denote existing observations (purple), possible detections (red) and faint sources (orange). The dashed line at 300~$\mu$Jy lies close to the limit of current detections. A total of 47 discs lie around or above this limit, representing a substantial potential addition to the REASONS sample of 74 systems \citep{2025Matra}.}
    \label{fig:band6_predictions}
\end{figure*}

In addition to the fitted dust parameters presented in the previous sections, we obtain predicted flux densities for the debris discs for those systems that have not yet been spatially resolved at (sub)millimetre wavelengths. We assumed the value of $q$ to be 3.5 for those discs without millimetre photometry, which coincidentally lies close to the peak of the observed distribution (as shown in Figure \ref{fig:q_hist}), such that these predictions should be relatively accurate. A summary of the Band 6 (1.3~mm) flux density predictions as a function of stellar luminosity and system age are presented in Figure \ref{fig:band6_predictions}. 

Amongst the 133 debris discs analysed in this work, 74 were part of the REASONS survey with high quality millimetre photometry \citep{2025Matra}. A further 20 had sub-millimetre photometry from another observatory \citep[e.g. JCMT,][]{2017Holland}, but were not observed at high angular resolution, and the remaining 39 systems lack any constraints on their millimetre brightness or architecture. 

We find that there exists an untapped population of far-infrared spatially resolved discs across the full range of stellar luminosities. These as-yet-unobserved systems are generally fainter ($F_{\rm disc} \leq 1~\rm{mJy}$) and found predominantly around older stars ($t_{\star} > 100~\rm{Myr}$). As such they occupy an under explored region of parameter space for both disc and disc-planet interaction. Based on observation with the ALMA 12-m array, we can define a disc as observable if its diameter can be spatially resolved by $\geq 3$ beams with a signal-to-noise $\geq 5$ per beam, equivalent to the REASONS sample. For a face-on disc with a diameter of 3 beams and an unresolved belt width, this requires an rms of 10~$\mu$Jy per beam, and an integrated flux density of 300~$\mu$Jy (equivalent to the horizontal dashed line in Figure \ref{fig:band6_predictions}).

Not all as-yet-undetected debris discs are amenable to characterisation by ALMA. Some lie too far north for observation (although the SMA or NOEMA could serve instead), are too extended on the sky, or their integrated flux densities are too low for high angular resolution observations. Under these constraints, a further 47 systems are potentially characterisable, representing an $\simeq50\%$ addition to the ensemble of millimetre-wavelength spatially resolved debris discs. The measurement of these additional discs at millimetre wavelengths would in almost all cases provide a first unique measurement of $q$ (and $M_{\rm dust}$) for those systems and substantially reduce the uncertainties on $s_{\rm min}$ (particularly its upper bound, see Table \ref{tab:app_results}). The improvement on the uncertainties of the calculated parameters for the ensemble would be $\simeq$ 20\%.

\subsubsection{Band 1 predictions}

\begin{figure}
    \centering
    \includegraphics[width=\columnwidth]{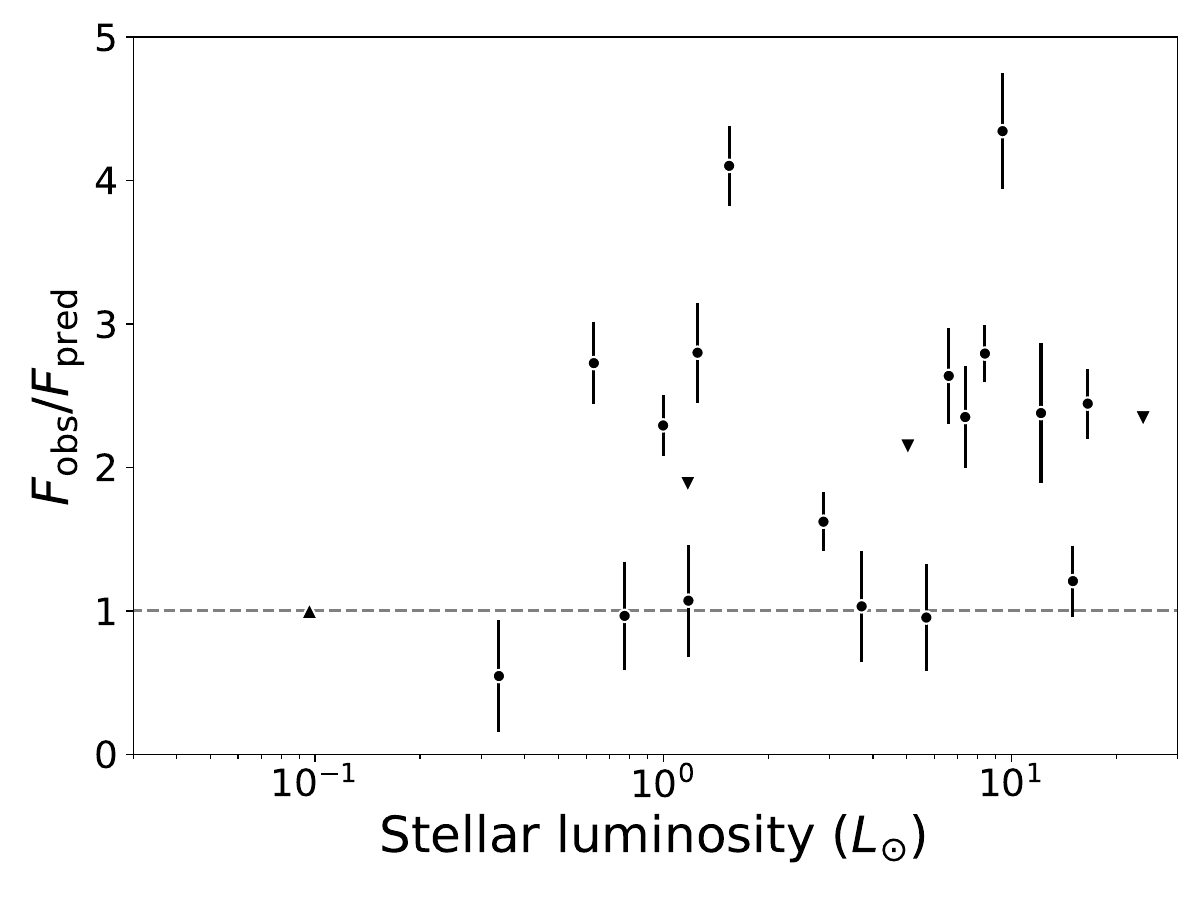}
    \caption{Plot of predicted and observed deep millimetre fluxes for systems in this work, separated by $L_{\star}$. The majority of detected discs have fluxes greater than the predictions made here indicating either a strong contribution from stellar chromospheric emission, or a flatter size distribution for the dust at millimetre to centimetre wavelengths. No system is observed to be significantly fainter than the predicted emission, such that determination of the detectability of a disc made from extrapolation should be considered reliable.}
    \label{fig:fobs_vs_fpred}
\end{figure}

We further calculated the SEDs of the debris discs to $\lambda_{\rm max} = $10~mm in order to compare the predicted and observed emission for the small population of debris discs which have ATCA or VLA measurements at 7 or 9~mm \citep{2012Ricci,2015Ricci,2016Macgregor,2017Marshall,2021Norfolk}. We find that for most detected systems the observed flux densities are a factor of two to three higher than the predicted emission from our radiative transfer modelling, a minority of the systems are consistent with our predictions, and none are significantly fainter. We plot the results of this comparison in Figure \ref{fig:fobs_vs_fpred}. The estimates for $\simeq~$cm wavelength emission from debris discs derived from fitting infrared to (sub-)millimetre photometry are therefore conservative in their predictions.

Stellar chromospheric activity can produce enhanced emission at millimetre to centimetre wavelengths above the level expected for simple stellar photosphere models \citep{2013Cranmer,2015Liseau,2018White}. Chromospheric emission for A type stars agrees with model predictions but the degree of excess chromospheric emission increases toward later spectral types \citep{2021Mohan}. The greatest amount of chromospheric excess is $\simeq$80\% of the predicted stellar photospheric model, such that this could not account for the disparity between the radiative transfer model predictions and measured flux densities. 

Recent modelling of discontinuous and power-law size distributions for dust grains in debris discs by \cite{2024KimWolf} provides an immediate explanation for the differences, as their work predicts overabundant small grains in a debris disc results in a flattening of the millimetre spectral slope compared to the sub-millimetre slope. Indeed, almost all systems for which we have enhanced long wavelength emission (compared to our prediction) are known to have haloes of small dust grains.

In addition to the already existing population of 7- or 9-mm detected discs \citep{2021Norfolk}, the modelling presented here points to a further 20 discs that would be detectable with ALMA Band 1 without the need for elevated millimetre-centimetre emission (i.e. $F_{\rm 7mm} \geq 15~\mu{\rm Jy}$, consistent with existing detections at 3 to 5$\sigma$). However, these systems would not be spatially resolved such that the relative contribution from the stellar chromosphere could not be independently constrained for a `clean' measurement of the millimetre-centimetre spectral index for these systems.

\section{Discussion}
\label{sec:dis}

All debris discs are assumed to consist of dust grains with a composition of astronomical silicate \citep{2003Draine}. This is primarily problematic on the basis that astronomical silicate is not a real material per se, despite its wide-found usage to fit dust emission. The secondary problem with this assumption is that there is good theoretical and observational evidence for water ice at the $\simeq10$\% level in debris disc dust grains \citep[e.g.][]{2012Gaspar,2016Morales,2018Marshall,2018Choquet}, and the recent direct detection of ice around HD~181327 \citep{2025Chen}. The presence of icy inclusions in silicate dust grains has two effects on the inferred dust parameters from our models; it increases the minimum grain size $s_{\rm min}$ and flattens the exponent of the size distribution $q$ \citep{2022Stuber}. However, not all debris discs need icy grains to fit their emission and the inferred fraction of ice required is also different for each system. Similar effects on $s_{\rm min}$ and $q$ are seen for enhanced porosity. Imaging polarimetry of debris discs suggest high porosity dust is required to match the observed polarisations at near-infrared wavelengths \citep{2020Esposito}, although this is not true of all systems \citep[e.g.][]{2020Marshall}. Since the true composition (and structure) of debris dust is largely unknown, and whatever biases this assumption introduces should be systematic, we expect that it would not affect the overall conclusions of this work.

As a simplifying assumption for the radiative transfer modelling the dust grain properties in the disc are universal with the same size distribution and minimum/maximum grain sizes at all radii within a given disc. From the results we clearly see that for a broad range of stellar luminosities that the minimum grain size is below the radiation blowout limit for the star. A substantial contribution to the total dust emission therefore comes from dust grains close to that limit. Such grains would be impacted by radiation forces driving them onto higher eccentricity orbits outside their point of formation within the debris belt, whereas larger grains would not be (as strongly) affected. Redistributing smaller grains to larger radii around a star lowers their temperature, such that reproducing the observed emission of the disc would require a smaller minimum grain size and flatter size distribution for the dust grains.

For the observations being modelled here we adopted a largest grain size for the dust $\simeq$ the longest wavelength of observation (i.e. $s_{\rm max} \simeq \lambda_{\rm max}$). However, larger grains up to cm sizes contribute a small fraction of the total dust emission at those wavelengths. We are therefore systematically underestimating the total emission at the longest wavelengths in our models. However, this effect is relatively small to begin with, being around 5\% of the total emission at 1.3~mm, and comparable in magnitude to the uncertainty on the observations. The impact of the absence of cm-sized grains from the model will also be smaller for the steeper size distributions, such that their inclusion would not serve to substantially reduce the trend of $q$ with $R_{\rm disc}$ identified in Figure \ref{fig:q_vs_rdisc}. 

We identified a significant slope to a power law fit between disc radius $R_{\rm disc}$ and $q$ within the sample of debris discs. The slope is consistent with dependence of $q$ on 1/$\sqrt{R_{\rm disc}}$, which may be interpreted as $q \propto v_{\rm Kep}$. The physical basis of this relationship is uncertain. We see no strong dependence between $q$ and $v_{\rm rel}$ by way of inferring the scale heights for vertically resolved discs which might point to its origin as dust size dependence on collision velocity (see Figure \ref{fig:q_vs_vrel}).

\section{Conclusions}
\label{sec:con}

We have modelled the spectral energy distributions for a sample of 133 spatially resolved debris disc systems. We calculated the minimum grain size, size distribution exponent, and dust masses for the sample. Based on the maximum likelihood models from the spectral energy distributions we further calculate predicted flux densities for these discs at 1.3~mm and 7~mm, analogous to ALMA bands 6 and 1, respectively.

Consistent with previous analyses of a smaller sample of 34 spatially resolved discs examined by \citet{2014Pawellek}, we recover the same trend in $L_{\star}$ vs. $s_{\rm min}$. With the larger sample examined here we find that the majority of discs around stars above 20~$L_{\odot}$ have values of $s_{\rm min}$ below the blowout limit, implying that a substantial population of grains small enough to be blown out by radiation pressure contributes to their observed far-infrared emission, consistent with the findings of \citet{2019ThebaultKral}.

We determine the median size distribution exponent to be $q = 3.49^{+0.38}_{-0.33}$. The majority of debris discs have size distribution exponents $q$ within the range 3 to 4, consistent with being dictated by the strength of the colliding bodies \citep{2012PanSchlichting}. A handful of outlier systems are found both above and below this range, albeit with uncertainties large enough to render them formally consistent with the expected range of $4 > q > 3$ for debris dust. We further identify a trend between $R_{\rm disc}$ and $q$ for these systems. Interpretation of this trend is uncertain, being attributable to either velocity dependent collisional fragmentation \citep{2012Gaspar} or grain growth in the outer regions of debris discs \citep{2024Kadono}. Additional high signal-to-noise, high resolution measurements of disc architectures at sub-millimetre wavelengths would help refine the relationship between the relative velocities of colliding bodies within these discs and the measured size distribution of the resulting dust grains.

We find no difference in the distributions of dust properties ($s_{\rm min}$, $q$) for those systems which are gas-bearing compared to the wider sample. The assumed belt architecture for the systems does not account for the differences in scattered light and millimetre wavelength morphology for the respective dust grain populations. Detailed imaging has demonstrated a diversity of morphologies exist for such systems (e.g. haloes of dust grains, radial separation between the scattered light and millimetre peak surface brightnesses). Overall we may conclude that the impact of the presence of gas on dust properties is therefore weak in general, although perhaps important in the specifics.

We have calculated the dust masses for debris discs within the sample and demonstrate that the measured dust masses are consistent with evolution from observed properties of protoplanetary discs and young debris discs in star forming regions \citep{2023Manara}. We convert the dust masses into disc masses for the sample following the three power law model presented in \citet{2021KrivovWyatt}, but substituting their assumed $q = 3.5$ for the $q$ measured in this analysis. The inferred disc masses do not reach the same high levels presented in that work such that, at least under the assumption of a 200~km maximum size for bodies in the disc, the total masses we calculate are consistent with the constraint of gravitational stability for protoplanetary discs \citep{2009Isella}.

Finally, we calculate flux densities for the sample extrapolated from the model fits at infrared to sub-millimetre wavelengths. At millimetre wavelengths, we identify a sub-set of 47 discs amenable to detection or more detailed characterisation at millimetre wavelengths. These discs are bright enough to be spatially resolved in a manner consistent with the existing REASONS sample \citep{2025Matra}. These systems are systematically fainter and older than the REASONS sample, and their characterisation would reveal a new region of parameter space for disc evolution studies. At centimetre wavelengths, we find that a the size distribution inferred at far-infrared and centimetre wavelengths are not necessarily in tension such that an excess of small grains or discontinuous size distribution is not always necessary to describe centimetre-bright debris discs \citep{2024KimWolf}. We identify a sub-set of 20 discs bright enough for detection at $\simeq$~centimetre wavelengths, the measurement of which would effectively double the number of such systems observed at SNR $\geq 5$ \citep{2021Norfolk}.

\section*{Acknowledgements}

The authors thank the anonymous referee for their detailed comments that helped improve the contents of the manuscript. JPM thanks Dr. Carlo Manara for sharing the code and data from \citet{2023Manara} which were used in making Figure \ref{fig:mdust_vs_age}.

This research has made use of the SIMBAD database, operated at CDS, Strasbourg, France \citep{2000Wenger}. This research has made use of the Astrophysics Data System, funded by NASA under Cooperative Agreement 80NSSC21M00561.

JPM acknowledges research support by the National Science and Technology Council of Taiwan under grant NSTC 112-2112-M-001-032-MY3.
RY acknowledges research support from the ASIAA Summer Student Program.
FK acknowledges support from the Spanish Ministry of Science, Innovation and Universities, under grant number PID2023-149918NB-I00. This work was also partly supported by the Spanish program Unidad de Excelencia María de Maeztu CEX2020-001058-M, financed by MCIN/AEI/10.13039/501100011033.
SZ acknowledges support from the Research Fellowship Program of the European Space Agency (ESA). 
LM acknowledges funding by the European Union through the E-BEANS ERC project (grant number 100117693). Views and opinions expressed are however those of the author(s) only and do not necessarily reflect those of the European Union or the European Research Council Executive Agency. Neither the European Union nor the granting authority can be held responsible for them.

\textit{Facilities:} None.

\textit{Software:} This paper has made use of the Python packages {\sc astropy} \citep{2013AstroPy,2018AstroPy}, {\sc SciPy} \citep{2020SciPy}, {\sc NumPy} \citep{2020Harris}, {\sc matplotlib} \citep{2007Hunter}, {\sc MiePython} \citep{MiePython}, {\sc emcee} \citep{2013ForemanMackey}, and {\sc corner} \citep{2016ForemanMackey}.

\section*{Data availability}

The data underlying this article and the scripts used in their analysis are available through the GitHub repository \hyperlink{https://github.com/jontymarshall/SEDs_Of_Spatially_Resolved_Discs}{https://github.com/jontymarshall/SEDs\_Of\_Spatially\_Resolved\_Discs}.



\bibliographystyle{mnras}
\bibliography{rdd_refs}


\appendix
\onecolumn
\section{Data Table}
\label{app:results}

Here we provide the source properties for the 133 debris discs with well-modelled SEDs along with the calculated values for their dust properties, namely $s_{\rm min}$, $q$, and $M_{\rm dust}$.

\begin{table*}
    \renewcommand{\arraystretch}{1.2}
    \centering
    \caption{Source parameters for the debris discs ($L_{\star}$, $R_{\rm disc}$, $\sigma_{\rm disc}$, $s_{\rm blow}$) and the dust properties derived from radiative transfer modelling ($s_{\rm min}$, $q$, $M_{\rm dust}$, $M_{\rm disc}$). Systems with a warm inner component to their disc are denoted by a $\dagger$ symbol next to the name. We also include predicted flux densities for $\lambda_{\rm obs} =$ 1.3~mm and 7~mm (ALMA B6 and B1), based on the maximum amplitude probability model. Potentially observable systems in B6, shown in Figure \ref{fig:band6_predictions}, have predicted flux densities consistent with $300~\mu$Jy whilst potentially observable systems in B1 have predicted flux densities consistent with $15~\mu$Jy. \label{tab:app_results}}
    \begin{tabular}{lccccccccccc}
    \hline
    Name &   Luminosity  & Mass & $R_{\rm disc}$ & FWHM & $s_{\rm blow}$ & $s_{\rm min}$ & $q$  & $M_{\rm dust}$         & $M_{\rm disc}$ & $F_{\rm B6, pred}$& $F_{\rm B1, pred}$\\
         & ($L_{\odot}$) & ($M_{\odot}$) & (au)  & (au)  &    ($\mu$m)    &   ($\mu$m)    &      & ($10^{-3} M_{\oplus}$) & ($M_{\oplus}$) & (mJy) & ($\mu$Jy) \\
    \hline\hline
HD216956C & 0.005 & 0.219 & 26 & $\leq $8 & --- & $10.88^{+2.77}_{-3.81}$ & $3.9^{+0.3}_{-0.2}$ & $1.9^{+0.3}_{-0.4}$ & $2.8^{+16.1}_{-7.1}$ & $0.263^{0.047}_{0.053}$ & $1.2^{0.2}_{0.2}$ \\
GJ581 & 0.012 & 0.283 & $25^{+3}_{-7}$ & $19^{+3}_{-8}$ & --- & $2.88^{+1.53}_{-1.51}$ & 3.5 & $0.7^{+0.1}_{-0.1}$ & $7.3^{+1.0}_{-0.7}$ & $0.276^{0.037}_{0.027}$ & $2.3^{0.2}_{0.2}$ \\
GJ3760 & 0.049 & 0.421 & $59^{+9}_{-15}$ & $34^{+14}_{-16}$ & --- & $7.50^{+1.34}_{-3.70}$ & 3.5 & $13.6^{+4.9}_{-1.8}$ & $148.9^{+53.4}_{-19.7}$ & $2.718^{0.974}_{0.359}$ & $17.6^{6.3}_{2.3}$ \\
HD197481 & 0.096 & 0.512 & $34^{+0}_{0}$ & $12^{+0}_{-0}$ & --- & $0.13^{+0.64}_{-0.10}$ & $3.1^{+0.1}_{-0.1}$ & $6.7^{+2.0}_{-1.4}$ & $138.0^{+51.3}_{-35.8}$ & $6.651^{1.991}_{1.416}$ & $60.6^{17.8}_{12.6}$ \\
GJ14 & 0.109 & 0.531 & $99^{+3}_{-8}$ & $33^{+7}_{-8}$ & --- & $7.86^{+2.16}_{-4.68}$ & $3.4^{+0.2}_{-0.2}$ & $1.1^{+0.1}_{-0.1}$ & $13.9^{+6.9}_{-5.1}$ & $1.854^{0.232}_{0.193}$ & $13.0^{1.6}_{1.3}$ \\
TWA7 & 0.127 & 0.555 & $90^{+20}_{-20}$ & $90^{+20}_{-20}$ & --- & $0.12^{+0.35}_{-0.09}$ & $3.6^{+0.0}_{-0.1}$ & $48.4^{+61.9}_{-22.8}$ & $435.8^{+678.5}_{-243.2}$ & $0.924^{1.176}_{0.432}$ & $5.8^{7.1}_{2.6}$ \\
TYC93404371 & 0.191 & 0.623 & $130^{+20}_{-60}$ & $100^{+40}_{-60}$ & --- & $0.64^{+2.78}_{-0.58}$ & $3.3^{+0.1}_{-0.2}$ & $13.7^{+9.3}_{-4.4}$ & $201.7^{+208.7}_{-107.8}$ & $3.625^{2.467}_{1.155}$ & $25.6^{17.3}_{8.1}$ \\
HD128311 & 0.288 & 0.700 & $50^{+5}_{-15}$ & $30^{+11}_{-15}$ & --- & $4.16^{+4.30}_{-3.21}$ & 3.5 & $0.3^{+0.2}_{-0.1}$ & $3.7^{+2.6}_{-0.8}$ & $0.169^{0.105}_{0.033}$ & $1.5^{0.6}_{0.2}$ \\
HD23356 & 0.298 & 0.708 & $60^{+5}_{-9}$ & $18^{+9}_{-9}$ & --- & $5.74^{+19.74}_{-4.88}$ & 3.5 & $0.2^{+0.1}_{-0.0}$ & $1.7^{+1.4}_{-0.4}$ & $0.163^{0.111}_{0.029}$ & $1.7^{0.6}_{0.2}$ \\
HD192263 & 0.304 & 0.711 & $85^{+4}_{-6}$ & $16^{+8}_{-6}$ & --- & $0.26^{+0.94}_{-0.23}$ & 3.5 & $1.0^{+1.7}_{-0.4}$ & $11.2^{+18.4}_{-4.3}$ & $0.170^{0.261}_{0.062}$ & $1.4^{1.6}_{0.4}$ \\
HD22049$^{\dagger}$ & 0.337 & 0.733 & $68^{+0}_{-1}$ & $12^{+1}_{-1}$ & --- & $0.51^{+3.28}_{-0.46}$ & $3.2^{+0.1}_{-0.1}$ & $0.6^{+0.3}_{-0.2}$ & $10.5^{+7.3}_{-4.0}$ & $15.176^{8.267}_{3.992}$ & $120.9^{58.4}_{28.2}$ \\
HD89452 & 0.366 & 0.750 & $69^{+7}_{-13}$ & $35^{+13}_{-13}$ & --- & $6.98^{+1.40}_{-3.13}$ & 3.5 & $4.7^{+1.0}_{-0.3}$ & $51.3^{+11.0}_{-3.4}$ & $0.928^{0.197}_{0.062}$ & $5.9^{1.2}_{0.4}$ \\
HD92945 & 0.367 & 0.751 & $96^{+1}_{-3}$ & $80^{+3}_{-3}$ & --- & $5.95^{+1.57}_{-1.96}$ & $3.8^{+0.1}_{-0.1}$ & $7.4^{+1.0}_{-0.6}$ & $29.1^{+26.3}_{-19.7}$ & $2.677^{0.353}_{0.230}$ & $12.0^{1.5}_{1.0}$ \\
HD17925 & 0.399 & 0.769 & $25^{+2}_{-4}$ & $8^{+4}_{-4}$ & --- & $0.21^{+6.89}_{-0.19}$ & $3.2^{+0.8}_{-0.3}$ & $0.4^{+0.5}_{-0.2}$ & $7.3^{+24.3}_{-4.2}$ & $0.269^{0.256}_{0.083}$ & $3.6^{2.0}_{0.7}$ \\
HD23484 & 0.408 & 0.774 & $49^{+4}_{-18}$ & $39^{+4}_{-18}$ & --- & $4.37^{+2.47}_{-2.05}$ & 3.5 & $1.1^{+0.2}_{-0.1}$ & $11.6^{+2.5}_{-1.5}$ & $0.618^{0.129}_{0.078}$ & $3.9^{0.7}_{0.4}$ \\
HD158633 & 0.415 & 0.778 & $33^{+6}_{-6}$ & $25^{+3}_{-6}$ & --- & $5.29^{+2.65}_{-3.01}$ & $4.1^{+0.6}_{-0.5}$ & $0.3^{+0.1}_{-0.1}$ & $0.7^{+3.8}_{-0.7}$ & $0.058^{0.010}_{0.005}$ & $1.2^{0.0}_{0.0}$ \\
HD131511 & 0.517 & 0.828 & $33^{+9}_{-5}$ & $9^{+5}_{-5}$ & --- & $7.25^{+38.91}_{-7.11}$ & 3.5 & $0.1^{+0.3}_{-0.0}$ & $0.7^{+3.0}_{-0.2}$ & $0.104^{0.207}_{0.017}$ & $2.4^{1.2}_{0.1}$ \\
HD31392 & 0.563 & 0.848 & $126^{+3}_{-24}$ & $104^{+8}_{-24}$ & --- & $2.35^{+0.94}_{-0.97}$ & 3.5 & $1.7^{+0.2}_{-0.2}$ & $18.2^{+2.5}_{-1.9}$ & $1.094^{0.149}_{0.115}$ & $6.4^{0.8}_{0.6}$ \\
HD53143 & 0.588 & 0.859 & $88^{+0}_{-2}$ & $23^{+2}_{-2}$ & --- & $0.66^{+0.98}_{-0.58}$ & $3.4^{+0.1}_{-0.1}$ & $3.5^{+2.1}_{-0.8}$ & $44.9^{+38.4}_{-16.9}$ & $1.867^{1.115}_{0.402}$ & $11.6^{6.5}_{2.4}$ \\
HD61005$^{\dagger}$ & 0.631 & 0.877 & $72^{+0}_{-1}$ & $38^{+1}_{-1}$ & --- & $4.10^{+3.06}_{-2.27}$ & $3.5^{+0.2}_{-0.2}$ & $23.2^{+6.1}_{-4.8}$ & $236.2^{+174.6}_{-125.5}$ & $3.860^{1.019}_{0.800}$ & $20.9^{5.5}_{4.3}$ \\
HD166 & 0.634 & 0.878 & $27^{+3}_{-3}$ & $14^{+5}_{-3}$ & --- & $2.49^{+2.48}_{-1.23}$ & $3.8^{+0.3}_{-0.3}$ & $0.9^{+0.2}_{-0.3}$ & $3.5^{+6.9}_{-4.6}$ & $0.106^{0.018}_{0.024}$ & $1.6^{0.1}_{0.1}$ \\
HD20794 & 0.646 & 0.883 & $32^{+3}_{-5}$ & $9^{+5}_{-5}$ & --- & $0.50^{+15.49}_{-0.47}$ & 3.5 & $0.1^{+0.2}_{-0.1}$ & $0.7^{+2.3}_{-0.6}$ & $0.231^{0.142}_{0.035}$ & $7.2^{0.7}_{0.2}$ \\
HD48370 & 0.774 & 0.929 & $88^{+7}_{-11}$ & $37^{+4}_{-12}$ & --- & $1.93^{+31.74}_{-1.72}$ & $2.7^{+0.5}_{-0.4}$ & $4.8^{+2.0}_{-1.3}$ & $133.7^{+131.6}_{-67.5}$ & $6.683^{2.748}_{1.783}$ & $72.5^{29.7}_{19.3}$ \\
HD115617 & 0.831 & 0.949 & $47^{+1}_{-4}$ & $36^{+4}_{-4}$ & --- & $0.37^{+3.96}_{-0.34}$ & $3.0^{+0.2}_{-0.1}$ & $0.3^{+0.2}_{-0.1}$ & $7.2^{+5.0}_{-2.6}$ & $1.980^{0.889}_{0.468}$ & $20.2^{7.5}_{3.9}$ \\
HD38858 & 0.843 & 0.952 & $110^{+10}_{-20}$ & $80^{+30}_{-20}$ & --- & $0.54^{+3.09}_{-0.50}$ & $3.3^{+0.1}_{-0.2}$ & $1.3^{+1.0}_{-0.4}$ & $19.3^{+20.5}_{-9.2}$ & $2.675^{2.167}_{0.759}$ & $17.8^{13.6}_{4.7}$ \\
HD30495 & 0.971 & 0.992 & $29^{+4}_{-4}$ & $22^{+3}_{-5}$ & --- & $3.78^{+2.61}_{-1.83}$ & 3.5 & $0.4^{+0.2}_{-0.1}$ & $4.9^{+1.7}_{-1.0}$ & $0.257^{0.070}_{0.042}$ & $3.1^{0.4}_{0.2}$ \\
HD202628 & 0.982 & 0.995 & $155^{+0}_{-3}$ & $16^{+3}_{-3}$ & --- & $6.36^{+1.16}_{-1.71}$ & $3.8^{+0.1}_{-0.1}$ & $1.2^{+0.1}_{-0.1}$ & $6.5^{+3.1}_{-2.4}$ & $1.052^{0.099}_{0.063}$ & $5.5^{0.5}_{0.3}$ \\
HD107146$^{\dagger}$ & 0.998 & 1.000 & $107^{+0}_{0}$ & $110^{+1}_{-1}$ & --- & $2.68^{+1.45}_{-1.31}$ & $3.4^{+0.1}_{-0.1}$ & $11.4^{+1.8}_{-1.5}$ & $158.7^{+53.0}_{-40.0}$ & $11.341^{1.792}_{1.473}$ & $72.4^{11.4}_{9.3}$ \\
HD73350 & 1.009 & 1.002 & $100^{+5}_{-18}$ & $85^{+4}_{-18}$ & --- & $2.99^{+3.04}_{-1.35}$ & 3.5 & $1.7^{+0.4}_{-0.3}$ & $18.0^{+4.7}_{-3.8}$ & $1.230^{0.315}_{0.256}$ & $7.7^{1.8}_{1.5}$ \\
HD202206 & 1.075 & 1.021 & $83^{+13}_{-16}$ & $39^{+21}_{-17}$ & 0.38 & $6.27^{+1.97}_{-2.19}$ & 3.5 & $1.2^{+0.2}_{-0.1}$ & $12.7^{+1.9}_{-1.1}$ & $0.248^{0.036}_{0.021}$ & $1.6^{0.2}_{0.1}$ \\
HD110897 & 1.116 & 1.032 & $63^{+5}_{-12}$ & $36^{+16}_{-13}$ & 0.40 & $2.99^{+3.17}_{-1.93}$ & 3.5 & $0.3^{+0.2}_{-0.1}$ & $3.6^{+1.9}_{-0.8}$ & $0.260^{0.120}_{0.053}$ & $2.5^{0.7}_{0.3}$ \\
HD377$^{\dagger}$ & 1.175 & 1.047 & $80^{+10}_{-10}$ & $20^{+20}_{-10}$ & 0.43 & $11.04^{+39.94}_{-10.02}$ & $3.5^{+0.8}_{-1.5}$ & $3.7^{+2.3}_{-1.3}$ & $39.0^{+128.9}_{-92.0}$ & $1.332^{0.832}_{0.460}$ & $6.9^{4.2}_{2.3}$ \\
HD45184 & 1.176 & 1.047 & $77^{+5}_{-31}$ & $64^{+3}_{-32}$ & 0.43 & $2.27^{+1.49}_{-0.99}$ & 3.5 & $1.1^{+0.2}_{-0.2}$ & $11.5^{+2.4}_{-2.0}$ & $0.595^{0.121}_{0.101}$ & $3.9^{0.6}_{0.5}$ \\
HD104860 & 1.180 & 1.048 & 110 & $\leq $200 & 0.43 & $4.21^{+3.93}_{-2.69}$ & $3.5^{+0.2}_{-0.2}$ & $6.7^{+2.5}_{-1.5}$ & $67.3^{+63.1}_{-37.6}$ & $2.714^{1.019}_{0.591}$ & $15.0^{5.6}_{3.2}$ \\
HD207129 & 1.218 & 1.058 & $150^{+2}_{-6}$ & $39^{+7}_{-6}$ & 0.45 & $1.39^{+1.52}_{-1.09}$ & $3.6^{+0.1}_{-0.1}$ & $1.5^{+0.6}_{-0.3}$ & $14.0^{+10.7}_{-6.1}$ & $2.146^{0.799}_{0.393}$ & $13.1^{4.3}_{2.1}$ \\
HD187897 & 1.231 & 1.061 & $90^{+10}_{-9}$ & $27^{+33}_{-10}$ & 0.45 & $2.94^{+0.94}_{-0.86}$ & 3.5 & $1.0^{+0.1}_{-0.1}$ & $10.4^{+1.1}_{-1.0}$ & $0.316^{0.033}_{0.031}$ & $2.1^{0.2}_{0.2}$ \\
HD1461 & 1.239 & 1.063 & $88^{+5}_{-8}$ & $28^{+27}_{-9}$ & 0.46 & $3.79^{+2.18}_{-1.29}$ & 3.5 & $0.5^{+0.1}_{-0.1}$ & $5.2^{+1.0}_{-0.9}$ & $0.347^{0.061}_{0.057}$ & $2.7^{0.3}_{0.3}$ \\
HD105 & 1.254 & 1.067 & 85 & $\leq $30 & 0.47 & $3.42^{+2.43}_{-1.26}$ & $3.2^{+0.1}_{-0.1}$ & $3.4^{+0.3}_{-0.3}$ & $59.8^{+13.7}_{-12.2}$ & $1.804^{0.172}_{0.176}$ & $15.0^{1.4}_{1.4}$ \\
HD38397 & 1.363 & 1.093 & $94^{+14}_{-7}$ & $34^{+18}_{-7}$ & 0.52 & $10.16^{+39.82}_{-7.23}$ & 3.5 & $5.8^{+2.5}_{-1.3}$ & $63.1^{+27.7}_{-14.1}$ & $1.376^{0.602}_{0.307}$ & $8.0^{3.4}_{1.7}$ \\
HD50554 & 1.499 & 1.123 & $75^{+5}_{-12}$ & $30^{+14}_{-13}$ & 0.59 & $2.12^{+1.05}_{-0.95}$ & 3.5 & $0.6^{+0.2}_{-0.1}$ & $6.8^{+1.7}_{-1.0}$ & $0.161^{0.038}_{0.023}$ & $1.3^{0.2}_{0.1}$ \\
HD82943 & 1.531 & 1.130 & $89^{+6}_{-21}$ & $62^{+14}_{-22}$ & 0.60 & $3.78^{+2.52}_{-1.41}$ & 3.5 & $1.3^{+0.2}_{-0.2}$ & $13.7^{+2.4}_{-2.0}$ & $0.750^{0.127}_{0.106}$ & $4.8^{0.7}_{0.6}$ \\
HD10647$^{\dagger}$ & 1.545 & 1.132 & $100^{+2}_{-6}$ & $70^{+5}_{-6}$ & 0.61 & $2.90^{+2.58}_{-1.42}$ & $3.6^{+0.1}_{-0.1}$ & $3.3^{+0.8}_{-0.7}$ & $32.0^{+19.2}_{-14.5}$ & $4.164^{1.042}_{0.812}$ & $23.4^{5.5}_{4.3}$ \\
HD48682 & 1.857 & 1.194 & $93^{+3}_{-12}$ & $83^{+1}_{-12}$ & 0.77 & $0.35^{+2.50}_{-0.31}$ & $3.3^{+0.1}_{-0.1}$ & $1.3^{+1.2}_{-0.6}$ & $19.3^{+22.4}_{-10.8}$ & $1.873^{1.602}_{0.800}$ & $14.3^{10.8}_{5.4}$ \\
HD219482 & 1.907 & 1.203 & $44^{+4}_{-6}$ & $12^{+6}_{-6}$ & 0.80 & $1.30^{+1.68}_{-0.80}$ & 3.5 & $0.5^{+0.2}_{-0.1}$ & $4.9^{+2.5}_{-1.6}$ & $0.130^{0.048}_{0.030}$ & $1.8^{0.2}_{0.2}$ \\
HD8907 & 1.972 & 1.214 & $100^{+3}_{-22}$ & $49^{+8}_{-22}$ & 0.83 & $4.77^{+1.69}_{-1.05}$ & $3.4^{+0.1}_{-0.1}$ & $3.0^{+0.2}_{-0.2}$ & $39.9^{+7.3}_{-7.2}$ & $2.192^{0.140}_{0.151}$ & $13.0^{0.8}_{0.9}$ \\
HD165908 & 2.144 & 1.243 & $130^{+5}_{-15}$ & $113^{+3}_{-15}$ & 0.93 & $7.60^{+8.46}_{-7.38}$ & $3.6^{+0.4}_{-1.2}$ & $0.2^{+0.1}_{-0.0}$ & $1.7^{+3.8}_{-4.2}$ & $0.884^{0.517}_{0.130}$ & $7.2^{2.8}_{0.7}$ \\
HD52265 & 2.263 & 1.263 & $124^{+2}_{-16}$ & $85^{+14}_{-16}$ & 1.0 & $0.81^{+0.90}_{-0.71}$ & 3.5 & $0.4^{+0.5}_{-0.1}$ & $4.0^{+5.5}_{-1.0}$ & $0.166^{0.202}_{0.037}$ & $1.6^{1.2}_{0.2}$ \\
HD35841 & 2.363 & 1.279 & $57^{+3}_{-8}$ & $15^{+9}_{-8}$ & 1.1 & $2.59^{+11.67}_{-2.36}$ & $3.1^{+0.4}_{-0.5}$ & $17.3^{+5.9}_{-4.6}$ & $321.2^{+290.0}_{-214.0}$ & $0.917^{0.314}_{0.243}$ & $6.4^{2.2}_{1.7}$ \\
HD111520 & 2.654 & 1.322 & $76^{+6}_{-30}$ & $50^{+20}_{-30}$ & 1.2 & $0.47^{+0.57}_{-0.42}$ & $3.2^{+0.1}_{-0.1}$ & $28.3^{+6.6}_{-3.1}$ & $515.0^{+181.0}_{-110.6}$ & $1.469^{0.343}_{0.161}$ & $10.3^{2.4}_{1.1}$ \\
HD146181 & 2.699 & 1.328 & 74 & $\leq $90 & 1.2 & $1.18^{+0.85}_{-0.99}$ & $3.3^{+0.2}_{-0.2}$ & $26.8^{+7.0}_{-4.6}$ & $394.2^{+217.5}_{-141.8}$ & $0.935^{0.245}_{0.159}$ & $5.9^{1.5}_{1.0}$ \\
HD191089 & 2.740 & 1.334 & $44^{+0}_{-3}$ & $16^{+3}_{-3}$ & 1.3 & $0.96^{+0.43}_{-0.50}$ & $3.2^{+0.1}_{-0.1}$ & $18.0^{+1.7}_{-1.0}$ & $316.8^{+61.4}_{-43.9}$ & $2.134^{0.197}_{0.116}$ & $16.2^{1.5}_{0.9}$ \\
    \hline
    \end{tabular}
\end{table*}

\begin{table*}
    \renewcommand{\arraystretch}{1.2}
    \centering
    \contcaption{}
    \begin{tabular}{lccccccccccc}
    \hline
    Name &   Luminosity  & Mass & Radius & FWHM & $s_{\rm blow}$ & $s_{\rm min}$ & $q$  & $M_{\rm dust}$         & $M_{\rm disc}$ & $F_{\rm B6, pred}$& $F_{\rm B1, pred}$\\
         & ($L_{\odot}$) & ($M_{\odot}$) & (au)  & (au)  &    ($\mu$m)    &   ($\mu$m)    &      & ($10^{-3} M_{\oplus}$) & ($M_{\oplus}$) & (mJy) & ($\mu$Jy) \\
    \hline\hline
HD189002 & 2.806 & 1.343 & $592^{+11}_{-136}$ & $469^{+38}_{-136}$ & 1.3 & $0.58^{+4.69}_{-0.53}$ & 3.5 & $11.2^{+28.8}_{-6.7}$ & $122.1^{+314.3}_{-73.3}$ & $1.600^{4.116}_{0.960}$ & $9.2^{23.5}_{5.5}$ \\
HD206893 & 2.852 & 1.349 & $108^{+4}_{-10}$ & $100^{+10}_{-10}$ & 1.3 & $6.42^{+2.52}_{-1.68}$ & $3.6^{+0.1}_{-0.2}$ & $3.0^{+0.3}_{-0.2}$ & $25.7^{+10.0}_{-12.6}$ & $1.365^{0.121}_{0.101}$ & $7.6^{0.6}_{0.5}$ \\
HD181327 & 2.883 & 1.353 & $81^{+0}_{0}$ & $16^{+0}_{-1}$ & 1.4 & $1.04^{+0.89}_{-0.76}$ & $3.1^{+0.1}_{-0.1}$ & $30.6^{+5.3}_{-4.2}$ & $578.1^{+196.2}_{-163.5}$ & $11.336^{1.966}_{1.575}$ & $89.4^{15.5}_{12.4}$ \\
HD127821 & 2.949 & 1.362 & 120 & $\leq $300 & 1.4 & $5.69^{+2.21}_{-2.86}$ & $3.8^{+0.2}_{-0.1}$ & $2.2^{+0.5}_{-0.2}$ & $7.8^{+12.4}_{-7.0}$ & $1.772^{0.386}_{0.170}$ & $7.7^{1.5}_{0.7}$ \\
HD129590 & 2.973 & 1.365 & $79^{+4}_{-30}$ & $70^{+20}_{-30}$ & 1.4 & $1.15^{+0.52}_{-0.67}$ & $3.4^{+0.1}_{-0.1}$ & $75.0^{+14.8}_{-8.5}$ & $952.9^{+416.1}_{-258.5}$ & $1.473^{0.291}_{0.166}$ & $8.5^{1.7}_{1.0}$ \\
HD112810$^{\dagger}$ & 3.145 & 1.387 & 90 & $\leq $200 & 1.5 & $4.34^{+0.72}_{-0.53}$ & $3.6^{+0.1}_{-0.1}$ & $10.0^{+0.7}_{-0.7}$ & $92.4^{+21.6}_{-20.5}$ & $0.517^{0.038}_{0.035}$ & $2.6^{0.2}_{0.2}$ \\
HD22484 & 3.148 & 1.388 & $46^{+4}_{-8}$ & $15^{+8}_{-8}$ & 1.5 & $1.06^{+1.69}_{-0.94}$ & 3.5 & $0.2^{+0.4}_{-0.1}$ & $2.2^{+4.0}_{-0.8}$ & $0.239^{0.183}_{0.035}$ & $5.7^{1.0}_{0.2}$ \\
HD145560$^{\dagger}$ & 3.221 & 1.397 & $76^{+4}_{-20}$ & $50^{+20}_{-20}$ & 1.6 & $19.96^{+47.22}_{-18.49}$ & $3.5^{+0.8}_{-1.3}$ & $17.8^{+8.5}_{-6.3}$ & $188.9^{+530.6}_{-396.5}$ & $1.231^{0.591}_{0.439}$ & $6.4^{3.1}_{2.3}$ \\
HD164249 & 3.241 & 1.399 & $62^{+3}_{-10}$ & $20^{+10}_{-10}$ & 1.6 & $4.22^{+0.50}_{-0.53}$ & $3.6^{+0.1}_{-0.1}$ & $12.2^{+0.9}_{-1.1}$ & $105.5^{+38.5}_{-35.7}$ & $1.097^{0.081}_{0.096}$ & $5.3^{0.4}_{0.4}$ \\
HD146897 & 3.352 & 1.413 & 82 & $\leq $90 & 1.6 & $0.88^{+0.72}_{-0.74}$ & $3.5^{+0.1}_{-0.2}$ & $92.9^{+44.7}_{-15.6}$ & $966.8^{+884.9}_{-434.0}$ & $1.313^{0.631}_{0.220}$ & $7.3^{3.5}_{1.2}$ \\
HD50571 & 3.362 & 1.414 & $190^{+20}_{-30}$ & $160^{+30}_{-30}$ & 1.7 & $4.60^{+2.36}_{-2.08}$ & $3.6^{+0.1}_{-0.1}$ & $1.2^{+0.2}_{-0.1}$ & $10.9^{+5.9}_{-4.3}$ & $1.551^{0.265}_{0.173}$ & $7.9^{1.2}_{0.8}$ \\
HD139664 & 3.386 & 1.417 & $75^{+6}_{-10}$ & $60^{+10}_{-10}$ & 1.7 & $1.26^{+1.83}_{-1.07}$ & $3.4^{+0.2}_{-0.1}$ & $1.7^{+0.7}_{-0.5}$ & $22.2^{+18.2}_{-9.8}$ & $2.182^{0.908}_{0.566}$ & $14.5^{5.2}_{3.2}$ \\
HD205674 & 3.456 & 1.425 & $160^{+10}_{-20}$ & $120^{+20}_{-20}$ & 1.7 & $3.47^{+0.44}_{-0.27}$ & $3.6^{+0.1}_{-0.1}$ & $4.3^{+0.3}_{-0.3}$ & $37.5^{+7.8}_{-8.7}$ & $1.211^{0.071}_{0.074}$ & $7.0^{0.4}_{0.4}$ \\
HD216435 & 3.546 & 1.436 & $68^{+8}_{-16}$ & $46^{+11}_{-16}$ & 1.8 & $5.75^{+6.75}_{-4.53}$ & 3.5 & $0.2^{+0.2}_{-0.0}$ & $2.7^{+1.9}_{-0.5}$ & $0.132^{0.071}_{0.018}$ & $1.6^{0.4}_{0.1}$ \\
HD170773 & 3.579 & 1.439 & $194^{+2}_{-5}$ & $68^{+5}_{-5}$ & 1.8 & $5.74^{+1.39}_{-1.32}$ & $3.6^{+0.1}_{-0.1}$ & $5.8^{+0.4}_{-0.3}$ & $52.7^{+12.7}_{-12.2}$ & $6.288^{0.456}_{0.377}$ & $33.6^{2.4}_{2.0}$ \\
HD15115$^{\dagger}$ & 3.712 & 1.455 & $93^{+1}_{-7}$ & $21^{+6}_{-7}$ & 1.9 & $7.03^{+8.60}_{-3.97}$ & $3.4^{+0.2}_{-0.3}$ & $4.5^{+1.2}_{-0.9}$ & $61.1^{+41.7}_{-36.4}$ & $1.867^{0.480}_{0.362}$ & $12.6^{3.2}_{2.4}$ \\
HD114082 & 3.788 & 1.463 & 38 & $\leq $40 & 1.9 & $1.32^{+0.71}_{-0.84}$ & $3.5^{+0.2}_{-0.1}$ & $41.5^{+7.4}_{-4.8}$ & $496.9^{+262.2}_{-156.7}$ & $0.621^{0.110}_{0.072}$ & $3.5^{0.6}_{0.4}$ \\
HD142446$^{\dagger}$ & 3.869 & 1.472 & $110^{+30}_{-40}$ & $80^{+50}_{-40}$ & 2.0 & $18.31^{+37.13}_{-13.57}$ & $3.8^{+0.5}_{-1.3}$ & $7.2^{+1.8}_{-1.5}$ & $27.9^{+97.3}_{-171.5}$ & $0.378^{0.095}_{0.077}$ & $1.4^{0.3}_{0.3}$ \\
HD113337 & 4.113 & 1.498 & $79^{+6}_{-3}$ & $30^{+17}_{-4}$ & 2.1 & $10.00^{+1.92}_{-1.90}$ & 3.5 & $1.1^{+0.1}_{-0.0}$ & $11.5^{+0.7}_{-0.5}$ & $0.609^{0.034}_{0.027}$ & $4.0^{0.2}_{0.1}$ \\
HD15745$^{\dagger}$ & 4.205 & 1.507 & $65^{+6}_{-20}$ & $50^{+10}_{-20}$ & 2.2 & $14.15^{+44.22}_{-12.86}$ & $3.4^{+1.1}_{-1.5}$ & $6.4^{+2.1}_{-1.7}$ & $89.9^{+233.2}_{-172.8}$ & $1.264^{0.425}_{0.346}$ & $8.4^{2.8}_{2.3}$ \\
HD147137$^{\dagger}$ & 4.345 & 1.522 & $120^{+80}_{-50}$ & $100^{+60}_{-50}$ & 2.3 & $17.56^{+42.37}_{-10.95}$ & $3.7^{+0.5}_{-1.3}$ & $5.6^{+1.0}_{-0.7}$ & $32.6^{+75.4}_{-141.2}$ & $0.372^{0.063}_{0.048}$ & $1.6^{0.3}_{0.2}$ \\
HD113556$^{\dagger}$ & 4.386 & 1.526 & 110 & $\leq $400 & 2.3 & $3.62^{+5.03}_{-0.77}$ & $4.3^{+0.4}_{-1.1}$ & $7.1^{+1.9}_{-1.6}$ & $7.1^{+31.7}_{-7.1}$ & $0.136^{0.035}_{0.030}$ & $0.4^{0.1}_{0.1}$ \\
HD176894 & 4.550 & 1.542 & $316^{+19}_{-36}$ & $124^{+63}_{-36}$ & 2.4 & $14.06^{+44.50}_{-7.00}$ & 3.5 & $2.2^{+0.8}_{-0.5}$ & $23.9^{+8.3}_{-5.2}$ & $2.122^{0.734}_{0.461}$ & $13.1^{4.5}_{2.8}$ \\
HD115600 & 4.798 & 1.565 & 90 & $\leq $200 & 2.6 & $0.54^{+0.34}_{-0.36}$ & $3.8^{+0.1}_{-0.1}$ & $25.3^{+8.9}_{-3.2}$ & $109.5^{+140.7}_{-80.9}$ & $0.211^{0.073}_{0.026}$ & $0.8^{0.3}_{0.1}$ \\
HD109085$^{\dagger}$ & 5.042 & 1.588 & $153^{+2}_{-3}$ & $53^{+2}_{-3}$ & 2.8 & $1.87^{+34.69}_{-1.82}$ & $2.4^{+0.3}_{-0.4}$ & $0.2^{+0.0}_{-0.0}$ & $8.4^{+2.3}_{-2.4}$ & $6.462^{0.540}_{0.466}$ & $83.9^{6.8}_{5.9}$ \\
HD16743$^{\dagger}$ & 5.113 & 1.594 & $159^{+2}_{-9}$ & $80^{+8}_{-9}$ & 2.8 & $4.71^{+1.23}_{-0.93}$ & 3.7$^{+0.1}_{-0.1}$ & $4.0^{+0.5}_{-0.4}$ & $43.5^{+4.9}_{-4.7}$ & $1.757^{0.198}_{0.189}$ & $9.2^{1.0}_{1.0}$ \\
HD117214 & 5.572 & 1.634 & 42 & $\leq $50 & 3.2 & $0.55^{+1.56}_{-0.51}$ & $3.3^{+0.1}_{-0.2}$ & $32.5^{+17.3}_{-7.7}$ & $492.1^{+396.7}_{-228.8}$ & $0.714^{0.378}_{0.169}$ & $4.1^{2.1}_{0.9}$ \\
HD37594 & 5.646 & 1.640 & $131^{+9}_{-4}$ & $35^{+63}_{-4}$ & 3.2 & $5.22^{+2.13}_{-1.27}$ & 3.5 & $2.6^{+0.3}_{-0.3}$ & $28.0^{+3.5}_{-3.5}$ & $1.738^{0.217}_{0.217}$ & $9.1^{1.1}_{1.1}$ \\
HD218396$^{\dagger}$ & 5.695 & 1.644 & $290^{+10}_{-30}$ & $250^{+30}_{-30}$ & 3.3 & $5.63^{+4.13}_{-4.12}$ & $3.6^{+0.3}_{-0.3}$ & $3.5^{+0.9}_{-0.6}$ & $29.7^{+32.9}_{-25.1}$ & $6.679^{1.808}_{1.151}$ & $34.6^{9.2}_{5.9}$ \\
HD106906 & 6.578 & 1.713 & $100^{+10}_{-40}$ & $80^{+30}_{-40}$ & 3.9 & $1.25^{+0.49}_{-0.72}$ & $3.6^{+0.2}_{-0.1}$ & $15.0^{+2.5}_{-1.9}$ & $120.3^{+87.9}_{-49.5}$ & $0.377^{0.061}_{0.047}$ & $1.8^{0.3}_{0.2}$ \\
HD95086 & 6.605 & 1.715 & $206^{+2}_{-3}$ & $180^{+4}_{-3}$ & 3.9 & $1.81^{+0.70}_{-0.94}$ & $3.5^{+0.1}_{-0.2}$ & $15.2^{+3.7}_{-3.0}$ & $179.3^{+104.8}_{-75.4}$ & $3.740^{0.913}_{0.726}$ & $23.5^{5.7}_{4.5}$ \\
HD27290 & 6.671 & 1.720 & $80^{+4}_{-8}$ & $71^{+1}_{-8}$ & 4.0 & $0.60^{+13.15}_{-0.56}$ & $2.9^{+0.7}_{-0.3}$ & $0.3^{+0.1}_{-0.0}$ & $7.2^{+9.3}_{-2.7}$ & $1.842^{0.747}_{0.293}$ & $17.9^{6.4}_{2.5}$ \\
HD105211$^{\dagger}$ & 6.790 & 1.729 & $131^{+0}_{-5}$ & $23^{+4}_{-5}$ & 4.1 & $7.94^{+42.37}_{-4.99}$ & $3.6^{+0.2}_{-1.3}$ & $0.7^{+0.1}_{-0.1}$ & $6.6^{+4.6}_{-18.8}$ & $2.359^{0.404}_{0.303}$ & $14.5^{2.0}_{1.5}$ \\
HD38529 & 6.892 & 1.736 & $108^{+7}_{-23}$ & $54^{+25}_{-23}$ & 4.2 & $2.23^{+3.98}_{-1.83}$ & 3.5 & $0.4^{+0.4}_{-0.1}$ & $4.7^{+4.6}_{-1.5}$ & $0.190^{0.145}_{0.047}$ & $2.3^{0.8}_{0.2}$ \\
HD121191$^{\dagger}$ & 7.198 & 1.758 & $55^{+4}_{-11}$ & $54^{+8}_{-11}$ & 4.4 & $1.75^{+0.81}_{-1.17}$ & $3.5^{+0.2}_{-0.2}$ & $23.0^{+5.9}_{-4.9}$ & $242.8^{+203.6}_{-115.3}$ & $0.313^{0.080}_{0.066}$ & $1.8^{0.4}_{0.4}$ \\
HD10638 & 7.276 & 1.763 & 160 & $\leq $400 & 4.5 & $2.71^{+1.93}_{-0.51}$ & $3.8^{+0.1}_{-0.2}$ & $5.6^{+1.1}_{-0.8}$ & $27.1^{+20.1}_{-23.7}$ & $1.243^{0.233}_{0.175}$ & $5.3^{0.9}_{0.7}$ \\
HD32297$^{\dagger}$ & 7.366 & 1.769 & $122^{+0}_{-3}$ & $62^{+4}_{-3}$ & 4.5 & $1.21^{+0.67}_{-0.92}$ & $3.3^{+0.1}_{-0.1}$ & $62.7^{+13.3}_{-10.3}$ & $913.5^{+431.7}_{-304.2}$ & $3.890^{0.826}_{0.641}$ & $23.9^{5.1}_{3.9}$ \\
HD195627$^{\dagger}$ & 7.877 & 1.803 & $119^{+8}_{-46}$ & $107^{+0}_{-47}$ & 4.9 & $5.86^{+1.83}_{-2.27}$ & $3.4^{+0.2}_{-0.2}$ & $1.2^{+0.1}_{-0.1}$ & $15.2^{+6.0}_{-4.7}$ & $2.712^{0.287}_{0.183}$ & $18.3^{1.8}_{1.1}$ \\
HD84870 & 8.044 & 1.814 & $260^{+50}_{-60}$ & $260^{+60}_{-60}$ & 5.1 & $2.44^{+0.39}_{-0.38}$ & $3.6^{+0.1}_{-0.1}$ & $6.7^{+0.6}_{-0.6}$ & $64.4^{+17.8}_{-16.5}$ & $1.962^{0.168}_{0.165}$ & $9.7^{0.8}_{0.8}$ \\
HD110058 & 8.308 & 1.831 & 50 & $\leq $100 & 5.3 & $0.71^{+0.49}_{-0.54}$ & $3.4^{+0.1}_{-0.1}$ & $32.4^{+6.2}_{-3.3}$ & $415.6^{+181.6}_{-109.8}$ & $0.664^{0.128}_{0.067}$ & $3.8^{0.7}_{0.4}$ \\
HD39060$^{\dagger}$ & 8.391 & 1.836 & $105^{+1}_{-3}$ & $92^{+3}_{-3}$ & 5.4 & $1.07^{+0.46}_{-0.46}$ & $3.6^{+0.1}_{-0.1}$ & $24.4^{+4.5}_{-3.5}$ & $193.8^{+86.5}_{-72.4}$ & $17.904^{3.316}_{2.545}$ & $85.9^{15.5}_{11.9}$ \\
HD30422 & 8.480 & 1.842 & $103^{+15}_{-25}$ & $55^{+24}_{-26}$ & 5.4 & $2.71^{+0.69}_{-0.82}$ & 3.5 & $0.6^{+0.1}_{-0.1}$ & $6.4^{+0.8}_{-0.9}$ & $0.132^{0.016}_{0.017}$ & $1.0^{0.1}_{0.1}$ \\
HD95698 & 8.680 & 1.854 & $147^{+13}_{-11}$ & $60^{+33}_{-11}$ & 5.6 & $6.39^{+3.84}_{-2.67}$ & 3.5 & $1.1^{+0.2}_{-0.1}$ & $11.9^{+2.2}_{-1.4}$ & $0.610^{0.109}_{0.068}$ & $4.0^{0.6}_{0.4}$ \\
HD131835$^{\dagger}$ & 9.430 & 1.899 & $83^{+0}_{-4}$ & $87^{+4}_{-4}$ & 6.2 & $4.56^{+34.04}_{-2.71}$ & $3.3^{+0.3}_{-0.9}$ & $28.4^{+6.6}_{-6.0}$ & $410.0^{+324.9}_{-509.3}$ & $2.140^{0.499}_{0.452}$ & $12.2^{2.8}_{2.6}$ \\
HD76582 & 9.882 & 1.924 & $219^{+9}_{-20}$ & $210^{+20}_{-20}$ & 6.6 & $5.92^{+17.58}_{-3.85}$ & $3.6^{+0.3}_{-0.4}$ & $2.4^{+1.5}_{-0.9}$ & $18.1^{+35.0}_{-19.4}$ & $2.442^{1.580}_{0.888}$ & $12.7^{7.9}_{4.4}$ \\
HD159492$^{\dagger}$ & 10.589 & 1.963 & $80^{+9}_{-21}$ & $53^{+12}_{-22}$ & 7.2 & $13.22^{+2.74}_{-7.39}$ & $4.8^{+1.0}_{-0.8}$ & $0.4^{+0.1}_{-0.1}$ & $0.1^{+3.2}_{-0.1}$ & $0.036^{0.003}_{0.002}$ & $0.8^{0.0}_{0.0}$ \\
HD21997 & 10.850 & 1.976 & $94^{+3}_{-5}$ & $52^{+5}_{-5}$ & 7.5 & $10.52^{+5.30}_{-3.80}$ & $3.4^{+0.2}_{-0.2}$ & $6.5^{+0.5}_{-0.5}$ & $89.2^{+35.5}_{-35.6}$ & $2.381^{0.186}_{0.196}$ & $14.3^{1.1}_{1.2}$ \\
HD20320 & 11.725 & 2.020 & $83^{+7}_{-23}$ & $53^{+17}_{-23}$ & 8.2 & $15.85^{+43.96}_{-7.35}$ & 3.5 & $0.2^{+0.0}_{-0.0}$ & $2.3^{+0.4}_{-0.2}$ & $0.288^{0.039}_{0.025}$ & $2.8^{0.2}_{0.1}$ \\
HD142091 & 12.014 & 2.035 & $102^{+4}_{-30}$ & $66^{+21}_{-31}$ & 8.5 & $6.01^{+44.10}_{-5.83}$ & $3.4^{+0.2}_{-1.3}$ & $0.6^{+0.5}_{-0.1}$ & $6.9^{+12.2}_{-14.5}$ & $1.104^{0.820}_{0.150}$ & $13.5^{5.0}_{0.9}$ \\
HD131488$^{\dagger}$ & 12.162 & 2.042 & $92^{+1}_{-8}$ & $46^{+7}_{-8}$ & 8.6 & $0.77^{+17.10}_{-0.66}$ & $2.8^{+0.4}_{-0.5}$ & $18.5^{+9.5}_{-7.4}$ & $471.7^{+496.6}_{-308.0}$ & $2.618^{1.350}_{1.047}$ & $24.8^{12.8}_{9.9}$ \\
HD153053 & 12.333 & 2.050 & $164^{+10}_{-65}$ & $126^{+18}_{-66}$ & 8.8 & $5.34^{+36.18}_{-2.92}$ & 3.5 & $0.8^{+0.3}_{-0.2}$ & $8.6^{+3.6}_{-2.3}$ & $0.680^{0.275}_{0.174}$ & $3.9^{1.4}_{0.9}$ \\
    \hline
    \end{tabular}
\end{table*}

\begin{table*}
    \renewcommand{\arraystretch}{1.2}
    \centering
    \contcaption{}
    \begin{tabular}{lccccccccccc}
    \hline
    Name &   Luminosity  & Mass & Radius & FWHM & $s_{\rm blow}$ & $s_{\rm min}$ & $q$  & $M_{\rm dust}$         & $M_{\rm disc}$ & $F_{\rm B6, pred}$& $F_{\rm B1, pred}$\\
         & ($L_{\odot}$) & ($M_{\odot}$) & (au)  & (au)  &    ($\mu$m)    &   ($\mu$m)    &      & ($10^{-3} M_{\oplus}$) & ($M_{\oplus}$) & (mJy) & ($\mu$Jy) \\
    \hline\hline
HD138965$^{\dagger}$ & 13.520 & 2.104 & $163^{+16}_{-3}$ & $38^{+7}_{-3}$ & 9.9 & $4.71^{+0.37}_{-0.24}$ & 3.7$^{+0.1}_{-0.1}$ & $5.2^{+0.2}_{-0.3}$ & $57.1^{+1.9}_{-2.8}$ & $1.663^{0.056}_{0.080}$ & $8.4^{0.3}_{0.4}$ \\
HD102647 & 13.695 & 2.112 & $31^{+2}_{-15}$ & $24^{+5}_{-16}$ & 10 & $5.84^{+6.30}_{-2.93}$ & $4.0^{+0.3}_{-0.9}$ & $0.3^{+0.1}_{-0.1}$ & $0.3^{+2.9}_{-4.7}$ & $0.460^{0.058}_{0.035}$ & $12.0^{0.2}_{0.1}$ \\
HD121617 & 13.738 & 2.114 & $78^{+5}_{-10}$ & $60^{+10}_{-10}$ & 10 & $1.68^{+0.34}_{-0.33}$ & $3.5^{+0.1}_{-0.1}$ & $55.4^{+4.8}_{-4.7}$ & $661.4^{+165.7}_{-136.5}$ & $2.208^{0.193}_{0.185}$ & $12.7^{1.1}_{1.1}$ \\
HD15257 & 14.455 & 2.145 & $270^{+60}_{-110}$ & $220^{+100}_{-110}$ & 11 & $4.99^{+13.06}_{-2.94}$ & $3.8^{+0.2}_{-1.2}$ & $1.2^{+0.7}_{-0.2}$ & $4.6^{+11.4}_{-25.0}$ & $0.883^{0.495}_{0.174}$ & $4.2^{1.8}_{0.6}$ \\
HD183324$^{\dagger}$ & 14.659 & 2.154 & $130^{+20}_{-14}$ & $34^{+49}_{-14}$ & 11 & $20.08^{+47.34}_{-14.21}$ & 3.5 & $0.1^{+0.0}_{-0.0}$ & $1.3^{+0.2}_{-0.1}$ & $0.108^{0.013}_{0.011}$ & $0.9^{0.1}_{0.1}$ \\
HD70313 & 14.799 & 2.159 & $115^{+11}_{-14}$ & $42^{+24}_{-14}$ & 11 & $5.35^{+38.70}_{-1.70}$ & 3.5 & $1.0^{+0.2}_{-0.2}$ & $10.4^{+2.1}_{-2.3}$ & $0.479^{0.094}_{0.101}$ & $2.9^{0.5}_{0.5}$ \\
HD110411 & 14.951 & 2.166 & $94^{+6}_{-26}$ & $56^{+24}_{-27}$ & 11 & $2.89^{+0.49}_{-1.00}$ & $3.4^{+0.2}_{-0.2}$ & $0.7^{+0.1}_{-0.0}$ & $8.6^{+3.6}_{-2.8}$ & $0.466^{0.035}_{0.025}$ & $3.5^{0.2}_{0.1}$ \\
HD9672$^{\dagger}$ & 15.020 & 2.169 & $136^{+2}_{-7}$ & $147^{+8}_{-7}$ & 11 & $4.93^{+2.72}_{-1.21}$ & $3.6^{+0.1}_{-0.2}$ & $8.8^{+1.5}_{-1.4}$ & $81.6^{+42.0}_{-38.2}$ & $4.404^{0.729}_{0.698}$ & $20.7^{3.4}_{3.2}$ \\
HD12467 & 15.130 & 2.173 & $128^{+22}_{-22}$ & $64^{+28}_{-23}$ & 11 & $14.21^{+53.53}_{-7.03}$ & 3.5 & $1.3^{+0.1}_{-0.1}$ & $14.0^{+1.2}_{-1.2}$ & $0.703^{0.061}_{0.059}$ & $4.3^{0.3}_{0.3}$ \\
HD141378 & 15.375 & 2.183 & $126^{+10}_{-13}$ & $60^{+19}_{-14}$ & 12 & $4.01^{+0.47}_{-0.28}$ & 3.5 & $1.1^{+0.1}_{-0.1}$ & $12.0^{+0.6}_{-0.8}$ & $0.516^{0.025}_{0.032}$ & $3.3^{0.1}_{0.2}$ \\
HD31295$^{\dagger}$ & 15.555 & 2.190 & $105^{+5}_{-28}$ & $64^{+21}_{-28}$ & 12 & $35.86^{+24.66}_{-27.19}$ & $3.9^{+0.7}_{-1.2}$ & $0.7^{+0.1}_{-0.1}$ & $2.0^{+12.0}_{-17.2}$ & $1.211^{0.127}_{0.092}$ & $5.6^{0.5}_{0.3}$ \\
HD216956 & 16.565 & 2.230 & $140^{+0}_{0}$ & $16^{+0}_{-0}$ & 13 & $8.29^{+4.64}_{-5.28}$ & $3.5^{+0.1}_{-0.2}$ & $0.8^{+0.2}_{-0.1}$ & $8.2^{+5.2}_{-3.2}$ & $27.697^{8.323}_{3.067}$ & $163.6^{42.0}_{15.5}$ \\
HD125162 & 17.339 & 2.259 & $81^{+6}_{-8}$ & $18^{+2}_{-8}$ & 14 & $1.38^{+1.30}_{-1.17}$ & $3.1^{+0.1}_{-0.1}$ & $0.7^{+0.1}_{-0.1}$ & $13.2^{+4.3}_{-3.6}$ & $1.170^{0.162}_{0.146}$ & $10.4^{1.2}_{1.1}$ \\
HD17848 & 18.427 & 2.299 & $151^{+8}_{-41}$ & $101^{+29}_{-41}$ & 15 & $25.52^{+48.42}_{-14.42}$ & 3.5 & $0.5^{+0.1}_{-0.0}$ & $5.5^{+0.6}_{-0.5}$ & $0.979^{0.104}_{0.088}$ & $5.9^{0.6}_{0.5}$ \\
HD138813$^{\dagger}$ & 18.818 & 2.313 & $120^{+10}_{-20}$ & $130^{+20}_{-20}$ & 15 & $9.64^{+3.47}_{-3.49}$ & $3.3^{+0.1}_{-0.1}$ & $5.9^{+0.6}_{-0.4}$ & $89.2^{+22.3}_{-18.2}$ & $1.256^{0.131}_{0.094}$ & $8.2^{0.8}_{0.6}$ \\
HD158352$^{\dagger}$ & 21.474 & 2.402 & $270^{+20}_{-60}$ & $380^{+90}_{-60}$ & 18 & $0.99^{+9.86}_{-0.93}$ & $3.3^{+0.2}_{-0.3}$ & $1.0^{+0.6}_{-0.3}$ & $16.5^{+14.9}_{-10.1}$ & $2.207^{1.261}_{0.614}$ & $17.2^{9.5}_{4.6}$ \\
HD192425$^{\dagger}$ & 22.061 & 2.420 & $197^{+8}_{-40}$ & $162^{+11}_{-40}$ & 19 & $24.81^{+34.61}_{-10.79}$ & 3.5 & $0.4^{+0.0}_{-0.0}$ & $4.2^{+0.5}_{-0.4}$ & $1.420^{0.153}_{0.119}$ & $8.5^{0.8}_{0.6}$ \\
HD11413 & 22.674 & 2.439 & $175^{+23}_{-14}$ & $41^{+28}_{-15}$ & 19 & $31.17^{+31.52}_{-13.53}$ & 3.5 & $0.3^{+0.0}_{-0.0}$ & $3.4^{+0.3}_{-0.2}$ & $0.351^{0.026}_{0.024}$ & $2.3^{0.1}_{0.1}$ \\
HD54341 & 23.398 & 2.461 & $170^{+20}_{-50}$ & $190^{+110}_{-50}$ & 20 & $5.26^{+21.09}_{-0.97}$ & $3.6^{+0.2}_{-0.5}$ & $3.5^{+0.4}_{-0.7}$ & $27.1^{+18.7}_{-32.6}$ & $0.762^{0.095}_{0.163}$ & $3.4^{0.4}_{0.7}$ \\
HD109573 & 23.899 & 2.476 & $77^{+0}_{0}$ & $15^{+1}_{-1}$ & 21 & $2.45^{+0.40}_{-0.34}$ & $3.5^{+0.1}_{-0.1}$ & $58.3^{+4.8}_{-4.7}$ & $611.3^{+136.8}_{-129.3}$ & $5.744^{0.471}_{0.465}$ & $26.9^{2.2}_{2.2}$ \\
HD182681 & 26.538 & 2.552 & $143^{+4}_{-10}$ & $100^{+10}_{-10}$ & 24 & $2.34^{+1.61}_{-2.01}$ & $3.3^{+0.3}_{-0.2}$ & $3.0^{+0.6}_{-0.8}$ & $48.3^{+30.7}_{-20.7}$ & $1.925^{0.411}_{0.510}$ & $12.4^{2.6}_{3.2}$ \\
HD38206$^{\dagger}$ & 27.005 & 2.564 & $180^{+20}_{-40}$ & $120^{+30}_{-40}$ & 24 & $8.78^{+1.87}_{-2.15}$ & $3.5^{+0.1}_{-0.1}$ & $1.3^{+0.1}_{-0.1}$ & $14.3^{+3.6}_{-3.6}$ & $1.069^{0.047}_{0.044}$ & $5.8^{0.2}_{0.2}$ \\
HD14055 & 28.954 & 2.616 & $180^{+10}_{-20}$ & $160^{+30}_{-20}$ & 26 & $2.64^{+0.53}_{-0.58}$ & $3.4^{+0.1}_{-0.1}$ & $1.0^{+0.1}_{-0.1}$ & $12.2^{+2.7}_{-2.3}$ & $2.657^{0.228}_{0.237}$ & $16.2^{1.2}_{1.3}$ \\
HD161868$^{\dagger}$ & 29.647 & 2.634 & $124^{+6}_{-10}$ & $110^{+10}_{-10}$ & 27 & $14.25^{+13.75}_{-5.23}$ & $3.6^{+0.2}_{-0.3}$ & $0.7^{+0.1}_{-0.1}$ & $6.3^{+3.7}_{-5.5}$ & $2.186^{0.224}_{0.194}$ & $13.1^{1.1}_{1.0}$ \\
HD188228 & 32.152 & 2.695 & $130^{+10}_{-14}$ & $38^{+25}_{-15}$ & 30 & $32.39^{+37.47}_{-18.68}$ & 3.5 & $0.0^{+0.0}_{-0.0}$ & $0.5^{+0.1}_{-0.1}$ & $0.261^{0.031}_{0.024}$ & $3.0^{0.2}_{0.1}$ \\
HD142315 & 32.189 & 2.696 & 140 & $\leq $500 & 30 & $1.34^{+0.42}_{-0.49}$ & $4.0^{+0.2}_{-0.2}$ & $7.9^{+1.1}_{-1.0}$ & $2.6^{+37.4}_{-28.2}$ & $0.177^{0.025}_{0.023}$ & $0.7^{0.1}_{0.1}$ \\
HD10939$^{\dagger}$ & 34.812 & 2.757 & $148^{+8}_{-21}$ & $66^{+33}_{-21}$ & 33 & $20.53^{+16.55}_{-5.83}$ & 3.5 & $1.0^{+0.1}_{-0.1}$ & $11.0^{+0.9}_{-0.8}$ & $1.376^{0.111}_{0.094}$ & $7.8^{0.6}_{0.5}$ \\
HD172167 & 55.933 & 3.157 & $118^{+4}_{-7}$ & $67^{+9}_{-7}$ & 61 & $13.34^{+6.52}_{-7.17}$ & $3.7^{+0.2}_{-0.2}$ & $0.2^{+0.0}_{-0.0}$ & $1.5^{+1.5}_{-1.2}$ & $10.341^{1.667}_{0.792}$ & $106.1^{6.8}_{3.2}$ \\
HD13161 & 72.240 & 3.397 & $101^{+5}_{-41}$ & $88^{+2}_{-41}$ & 85 & $10.66^{+4.41}_{-5.36}$ & $3.3^{+0.2}_{-0.2}$ & $0.5^{+0.1}_{-0.0}$ & $6.5^{+2.7}_{-2.4}$ & $1.358^{0.134}_{0.089}$ & $13.2^{0.8}_{0.5}$ \\
HD139006 & 72.664 & 3.403 & $41^{+4}_{-2}$ & $14^{+2}_{-2}$ & 86 & $5.11^{+3.06}_{-2.17}$ & 3.5 & $0.3^{+0.1}_{-0.0}$ & $3.1^{+0.7}_{-0.5}$ & $0.469^{0.053}_{0.038}$ & $10.1^{0.2}_{0.2}$ \\    
    \hline
    \end{tabular}
\end{table*}

\bsp	
\label{lastpage}
\end{document}